\documentclass[10pt]{article}
\usepackage{jheppub}
\pdfoutput=1
\usepackage{indentfirst}
\usepackage{graphicx,wrapfig,float,slashed}
\graphicspath{{Plots/}}
\usepackage{amsmath,amssymb,dsfont,epsfig,graphicx,xcolor}
\usepackage{epstopdf}
\usepackage{ragged2e}
\usepackage{enumerate}
\usepackage{multirow}
\usepackage{amsmath}
\usepackage{lipsum}
\usepackage{epstopdf}
\usepackage{comment}
\usepackage{xcolor}
\usepackage{tabu}
\usepackage[toc]{appendix}
\usepackage{enumitem}
\usepackage[normalem]{ulem}
\usepackage[font=small,skip=1pt]{caption}
\allowdisplaybreaks
\usepackage{pifont}
\usepackage{subcaption}
\usepackage{cancel}
\usepackage{multicol}
\usepackage{geometry}
\usepackage{pifont}

\definecolor{nicered}{rgb}{0.7,0.1,0.1}
\definecolor{nicegreen}{rgb}{0.1,0.5,0.1}
\definecolor{violet}{rgb}{0.7,0.3,0.3}
\hypersetup{colorlinks,citecolor= nicegreen,linkcolor= nicered}

\newcommand{\nc}{\newcommand}
\nc{\non}{\nonumber}
\nc{\hc}{\hbox {H.c.}}
\nc{\noi}{\noindent}
\nc{\barx}{\bar{x}}
\nc{\pbarn}{\;\hbox {pb}}
\nc{\fbarn}{\;\hbox {fb}}

\nc{\hsp}{\hspace{0.5cm}}
\nc{\lsp}{\hspace{1cm}}
\nc{\Lsp}{\hspace{2cm}}
\nc{\LLsp}{\lsp\lsp}
\nc{\lra}{\longrightarrow}
\nc{\p}{\prime}
\nc{\sgn}{\text{sgn}}
\nc{\ph}{\varphi}
\nc{\op}{{\cal O}}
\nc{\eq}{\text{Eq.~}}

\nc{\beq}{\begin{equation}}  \nc{\eeq}{\end{equation}}
\nc{\bea}{\begin{eqnarray}}  \nc{\eea}{\end{eqnarray}}
\nc{\baa}{\begin{array}}     \nc{\eaa}{\end{array}}
\nc{\bit}{\begin{itemize}}   \nc{\eit}{\end{itemize}}
\nc{\ben}{\begin{enumerate}} \nc{\een}{\end{enumerate}}
\nc{\bce}{\begin{center}}    \nc{\ece}{\end{center}}
\nc{\bpm}{\begin{pmatrix}}   \nc{\epm}{\end{pmatrix}}
\nc{\bvt}{\begin{verbatim}}  \nc{\evt}{\end{verbatim}}

\title{Learning the latent structure of collider events}
\author[\,\triangleleft]{B.~M.~Dillon}
\author[\,\circ]{D.~A.~Faroughy}
\author[\,\triangleleft\,\triangleright]{J.~F.~Kamenik} 
\author[\,\diamond ]{and~M.~Szewc}

\affiliation[\triangleleft\,]{Jo\v zef Stefan Institute, Jamova 39, 1000 Ljubljana, Slovenia}
\affiliation[\circ\,]{Physik-Institut, Universit\"at Z\"urich, CH-8057, Switzerland}
\affiliation[\triangleright\,]{Faculty of Mathematics and Physics, University of Ljubljana,
 Jadranska 19, 1000 Ljubljana, Slovenia}
\affiliation[\diamond\,]{International Center for Advanced Studies (ICAS) and CONICET, UNSAM,
Campus Miguelete, 25 de Mayo y Francia, CP1650, San Martín, Buenos Aires, Argentina}

\emailAdd{barry.dillon@ijs.si}
\emailAdd{faroughy@physik.uzh.ch}
\emailAdd{jernej.kamenik@ijs.si}
\emailAdd{mszewc@unsam.edu.ar}

\abstract{We describe a technique to learn the underlying structure of collider events directly from the data, without having a particular theoretical model in mind.
It allows to infer aspects of the theoretical model that may have given rise to this structure, and can be used to cluster or classify the events for analysis purposes.
The unsupervised machine-learning technique is based on the probabilistic (Bayesian) generative model of Latent Dirichlet Allocation. We pair the model with an approximate inference algorithm called Variational Inference, which we then use to extract the latent probability distributions describing the learned underlying structure of collider events.
We provide a detailed systematic study of the technique using two example scenarios to learn the latent structure of di-jet event samples made up of QCD background events and either $t\bar{t}$ or hypothetical $W' \to (\phi \to WW) W$ signal events.
}

\date{\today} 

\begin{document}
\maketitle

%%%%%%%%%%%%%%%%%%%%%%%%%%%%%%%%%%%%%%%%%%%%%%%%%%
%
\section{Introduction}
%
%%%%%%%%%%%%%%%%%%%%%%%%%%%%%%%%%%%%%%%%%%%%%%%%%%

\noindent 
With the discovery of the the Higgs boson~\cite{higgs_disc1,higgs_disc2}, all the degrees of freedom that form our current consistent theoretical understanding of fundamental quantum interactions -- the standard model (SM) -- have been experimentally established. The theoretical and phenomenological program that enabled this fundamental scientific achievement has lasted for more than three decades starting when the dominant SM Higgs boson production and decay modes have been first identified and computed~\cite{Gunion:1989we,Dittmaier:2011ti,Dittmaier:2012vm}. It involved detailed theoretical calculations of both the eventual signal, but also the most relevant (and typically much more abundant) background processes from which the (relatively small) signal had to be painstakingly extracted with the use of advanced statistical methods~\cite{annurev_statistics_062713_085841, higgs_disc1,higgs_disc2}. 

Contrary to the hunt for the SM Higgs boson (and also other SM heavy resonances, like the top quark or the weak gauge bosons), whose properties, processes and signatures at high energy colliders were well predicted and understood before their discovery, our current quest for uncovering possible new physics (NP) degrees of freedom beyond those of the SM faces a much bigger challenge. Namely, there is no unique well established model of NP which would convincingly address all the known SM shortcomings and whose phenomenology could be precisely studied and targeted experimentally.  Instead there exist a plethora of NP proposals and possibilities. The few simplest, most elegant and thus most compelling possibilities have already been mostly excluded or pushed into fringe corners of their respective parameter spaces, while systematically exploring the phenomenology of the whole vast imaginable model space is clearly beyond current human capabilities. 

In the last few years, machine learning (ML) tools have opened new avenues in NP searches, see e.g. \cite{Nachman:2019dol,Larkoski:2017jix,Guest:2018yhq,Bollweg:2019skg} and references therein. The currently most widely used framework is that of Neural Networks (NNs) as efficient likelihood approximators trained on vast amounts of data. Since these supervised ML approaches commonly rely on theoretical predictions for both NP (signal) and  SM (background) training data sets (typically through Monte Carlo (MC) generators), their use in searches for a priori unknown new phenomena in LHC events is severely limited.

There have been recent advances in unsupervised or semi-supervised ML techniques designed to be able to separate signal and background events in mixed samples, and could therefore be run directly on experimental data without the need for pure MC training samples, see e.g. refs.~\cite{Metodiev:2017vrx, vonBuddenbrock:2019bzi,Collins:2019jip,Heimel:2018mkt,Nachman:2020lpy,Andreassen:2020nkr,Kasieczka:2018lwf, DeSimone:2018efk, Blance:2019ibf, Collins:2018epr,Metodiev:2018ftz,Komiske:2018vkc, Komiske:2018oaa, Dery:2017fap, Cohen:2017exh, Agashe:2018leo,Aguilar-Saavedra:2017rzt,Hajer:2018kqm,Farina:2018fyg,Cerri:2018anq,DAgnolo:2018cun,DAgnolo:2019vbw, Amram:2020ykb, Romao:2020ojy, Knapp:2020dde,Aad:2020cws}. They rely on categorizing and comparing datasets with different expected signal and background admixtures or identifying anomalous events inside large datasets. While these approaches ameliorate the model dependence of fully supervised ML, they are still potentially susceptible to correlated systematics (i.e. detector) effects and/or subject to large look-elsewhere effects. In addition, they generally work best when applied on very large datasets. Consequently their performance may suffer when looking for effects in tails of distributions.

Recently~\cite{Dillon:2019cqt}, we have proposed a new technique to classify jets and events {\it in situ} within a single mixed event sample, using tools developed in a branch of ML called generative statistical modeling, see e.g.~\cite{Bishop:998831}. Developed primarily to identify emergent themes in collections of documents, these models infer the hidden (or latent) structure of a document corpus using posterior Bayesian inference based on word and theme co-occurence~\cite{other_blei,Blei03latentdirichlet,Griffiths5228,Pritchard945,Hofmann99probabilisticlatent,Deerwester90indexingby,Nigam99textclassification}. Using the example of jet substructure observables based on the clustering history, we have shown how to construct statistical mixed membership models of jet substructure. In particular, using the model of Latent Dirichlet Allocation (LDA)~\cite{Blei03latentdirichlet}, which can be solved efficiently using e.g. Variational Inference (VI)~\cite{Blei_2017} techniques, we were able to define robust parametric jet and event classifiers.\footnote{Some related ML techniques have been previously studied in Refs.~\cite{Agashe:2018leo,Agashe:2017wss,Agashe:2016rle,Collins:2018epr,deFavereau:2013fsa,Plehn:2009rk,Butterworth:2008iy,Cui:2010km,Kaplan:2008ie}.} 

In the present work we provide further details of this approach, building upon the basic assumptions and premises about the relevant measurements/observables and their statistical modeling, in order to construct generative Bayesian models most relevant and practical for particle physics event classification, in particular LDA. We also provide further justification for why jet clustering history observables are a particularly interesting and applicable example for these methods.  
Finally, we perform a systematic study of the parametric and Bayesian prior dependence and performance of VI and LDA, respectively,  based on two representative examples of boosted $t\bar t$ events and events containing hypothetical boosted color neutral, but hadronically decaying  scalars~\cite{Agashe:2018leo,Collins:2018epr,Collins:2019jip}. In particular, we identify a robust measure of LDA and VI performance, which does not rely on access to labelled data but at the same time correlates strongly with traditional classification performance measures (like tagging efficiency and mistag rate), and use it to identify parameter and prior ranges most suitable for the example datasets.

The paper is structured as follows: In Sec.~\ref{sec:probmodelling} we outline the statistical premises and introduce Bayesian generative models upon which LDA is based. We also provide details of LDA training and inference methods and how they can be applied to event classification. In Sec.~\ref{sec:latentjetsubstructure} we apply the general framework to the multi-jet event data in the form of jet clustering history observables and discuss the most appropriate data representations. 
{The benchmark event samples used for our study are introduced in Sec.~\ref{sec:sbm}, where we also discuss the data preparation steps that need to be considered when using LDA. }
Sec.~\ref{sec:ldaresults} contains the main results of our systematic study of LDA based classification methods applied to the example datasets.  Finally we summarize our conclusions and provide an outlook in Sec.~\ref{sec:conclusions}. 

%%%%%%%%%%%%%%%%%%%%%%%%%%%%%%%%%%%%%%%%%%%%%%%%%%
%
\section{Probabilistic generative modelling for collider experiments}
\label{sec:probmodelling}
%
%%%%%%%%%%%%%%%%%%%%%%%%%%%%%%%%%%%%%%%%%%%%%%%%%%

\noindent The goal in high energy collider experiments is to gain understanding of the underlying physical processes taking place at very high energies during the events, i.e. the hard collisions.
Each event can result in anywhere from $\mathcal{O}(10)$ to $\mathcal{O}(1000)$ particles being detected away from the beamline and the detector records the energy, momentum, and tracking information of these particles.
One must then analyse this high-dimensional dataset and compare it to what is expected from theoretical predictions in order to gain an understanding of the underlying physics.
To perform this analysis in practice, typically the dimensionality of the dataset must be drastically reduced.
How this is done depends on the type of underlying physics one wants to study and typically involves some combination of jet clustering and grooming, pile-up subtraction, the use of certain high-level observables, and performing cuts to remove unnecessary elements of the dataset.
Once the dataset has been processed and the relevant high-level observables have been collated, a statistical analysis comparing the measurements with the theoretical predictions can quantify how much a particular physics model agrees with the data.

Bayesian probabilistic generative modelling is an unsupervised machine learning approach in which one constructs a probabilistic model for a dataset, and then uses approximate inference techniques to estimate the parameters of this probabilistic model directly from the data.
If the probabilistic model is a good approximation to how the data was actually generated, this in turn allows to identify patterns in the dataset. 
In our case these patterns could contain important information on the underlying physical processes registered in collider events.

In the most general sense, a single event $e_j$ $(j=1,2,\ldots,N_e)$ can be represented by a finite list of measurements, $e_j=\{o_{j,1},o_{j,2},\ldots\}$, where $o_{j,i}$ $(i=1,2,\ldots,N_j)$ 
{are in general functions (or mappings) of the relevant multi-particle phase-space}.  
We can construct a model for the events by supposing that the measurements have been sampled from a (presumably complicated) joint probability distribution $p(e_j)=p(o_{j,1},o_{j,2},\ldots)$. 
This is the starting point for the unsupervised analysis techniques used in this paper. Writing a general statistical model describing the generative process of events is not possible in practice. To proceed, it is necessary to impose a set of simplifying assumptions on the joint probability distribution. The functional dependence of this distribution on $o_{j,i}$ must of course be flexible enough in order to account for the multiple physical processes manifest in each event, but it also must be simple enough such that efficient inference techniques can be implemented. In order to model the events using the techniques described in this paper, 
{the phase-space observables furthermore need to be labeled and binned} so that the possible measurements $o_{j,i}$ are discrete and finite in number. This requirement allows to construct probabilistic models based on multinomial distributions that describe the occurrence of the measurement bins $o_{j,i}$ in events. If we were to consider unbinned observables then $p(e_j)$ would be constructed from continuous probability distributions. However, given that in practice the measurements we work with are also coarse-grained due to detector granularity and reconstruction uncertainties it is intuitive to select bins for the measurements that reflect this.

%%%%%%%%%%%%%%%%%%%%%%%%%%%%%%%%%%%%%%%%%%%%%%%%%%
\subsection{Probabilistic generative models}\label{sec:models}
%%%%%%%%%%%%%%%%%%%%%%%%%%%%%%%%%%%%%%%%%%%%%%%%%%

\noindent  In order to introduce the reader to the concepts and models used in this work we will discuss two different models for $p(e_j)$ of increasing complexity: mixture models, and mixed-membership models.
The starting point for the construction of these models is de Finetti's representation theorem~\cite{Hewitt}, which states that if the measurements in the joint probability distribution are exchangeable then the measurements are conditionally independent given some latent variables.
The exchangeability requirement simply means that the joint probability distribution should be invariant under a re-ordering or exchanging of the measurements. 
Note that measurement exchangeability is not to be confused with measurement independence (i.i.d). The former is a weaker condition that leads to more flexible probabilistic models capable of capturing complicated hierarchical patterns in the data. 
Formally, the theorem states that the joint probability distribution can be written in the form
\beq\label{definetti}
p(o_{j,1},o_{j,2},...,o_{j,n})\ = \int_{\Theta} d\theta\, p(\theta)\prod_{i=1}^{N_j}p(o_{j,i}|\,\theta)\,,
\eeq
where $\theta$ ($\Theta$) represents a hidden latent parameter (space) which is marginalised over in the joint probability.
From the $p(o_{j,i}|\theta)$ factor on the right-hand side of the above equation we can see that the independence of the different measurements is manifest, and is replaced by a conditional dependence on the latent space variables. When viewing this result from a Bayesian perspective, the probability functions $p(o\,|\theta)$ represent likelihoods while the function $p(\theta)$ outside of the product acts as a prior distribution over the latent space. This theorem underpins probabilistic models such as mixture models and mixed-membership models, which we will discuss in the following sections.

\subsubsection{Mixture models}
%%%%%%%%%%%%%%%%%%%%%%%%%%%%%%%%%%%%%%%%%%%%%%%%%%

\noindent We now present one of the simplest probabilistic models for a sample of collider events. 
The mixture model consists of $T$ probability distributions over the measurement bins, represented by $p(o|t,\beta)$ for $t=1,\ldots,T$.
These probability distributions are $M$-dimensional multinomials (multidimensional generalisations of the binomial), where $M$ is the number of {bins in observable-space}. 
The parameters of these multinomials are represented by the elements of a $T\!\times\! M$ dimensional matrix, $\beta_{t,m}$ having the property that $\sum_{m=1}^M \beta_{t,m}=1$ for all $t$.
Each row in $\beta_{t,m}$ contains the parameters of the multinomial associated to one of the $T$ probability distributions.
A key feature of mixture models is that they assume measurements in a single event have been sampled from just one of these $T$ multinomial distributions.
So for each event one of the $T$ distributions is selected from a multinomial probability distribution $p(t|\omega)$, where $\omega=(\omega_1,\ldots,\omega_T)$ are the probabilities to select each $T$. They satisfy $0\le\omega_t\le1$ and $\sum_t^T\omega_t=1$. 
In this work we refer to each $T$ latent multinomial distribution as a `theme', in reference to the field of `topic modelling' in text analysis where these methods were popularised, thus the $\omega$ parameters are referred to as theme weights.

It is useful to describe the probabilistic model in terms of a generative process, outlining the underlying assumption on how the events were generated.
The generative process for a collection of events in a mixture model goes as follows:
\begin{enumerate}[label=\roman*.]
\itemsep0em
\item Randomly sample a theme $t\sim p(t|\omega)$. 
\item Randomly sample a measurement $o_{j,i}\sim p(o|t,\beta_{t})$. 
\item Repeat step (ii) for each measurement in the event.
\item Repeat steps (i)-(iii) for each event in the sample.
\end{enumerate}
The mathematical structure of the model can be realised by taking the representation of the joint probability distribution due to de Finetti's theorem \eqref{definetti} and defining $\theta\in\mathbb{R}$ with a prior distribution over the latent space as $p(\theta)=\sum_t^T p(\theta|\omega)\delta(\theta-t)$ where $p(\theta|\omega)$ is a density such that $p(\theta\!=\!t|\omega)=\omega_t$. 
This leads us to the form
\beq\label{eq:mixture}
p(e_j)=\sum_{t=1}^Tp(t|\omega) \prod_{i=1}^{N_j}\,p(o_{j,i}|t,\beta)\,.
\eeq

\begin{figure}[t!]
\begin{center}
\includegraphics[width=0.6\linewidth]{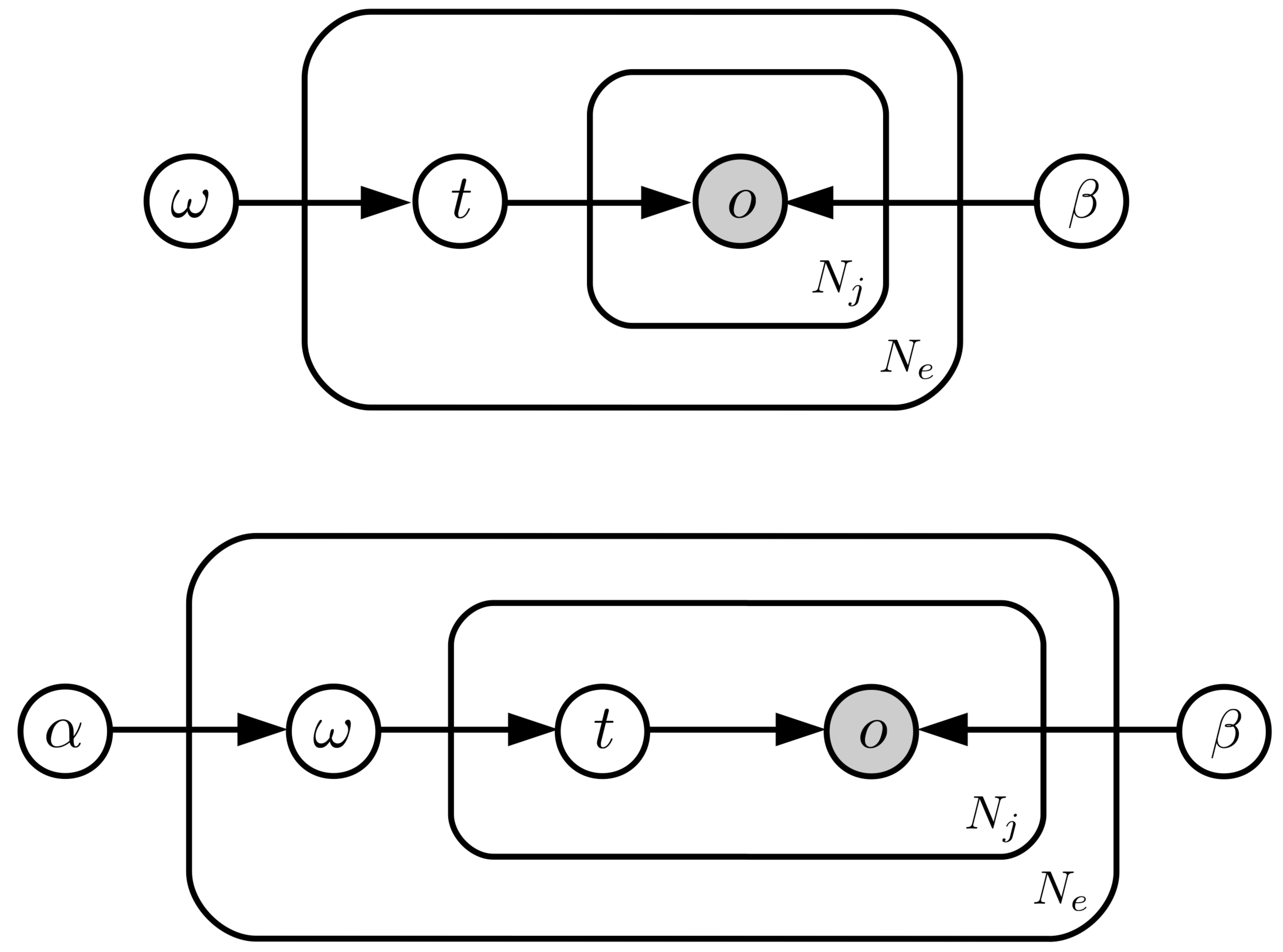}
\vspace{2pt}
\caption{The graphical models representing the generative process for $N_j$ measurements in $N_e$ events for the the mixture model Eq.~\eqref{eq:mixture} (upper diagram) and the mixed-membership model Eq.~\eqref{eq:lda} (lower diagram). See text for details. \label{graph_models}}
\end{center}
\end{figure}

The generative process described here can be visually represented using a so-called graphical model, see e.g.~\cite{Bishop:998831}: the unobserved variables (Latent random variables and model parameters) are represented by white circles, observed data (measurements) are represented by shaded circles, while the conditional dependencies and {i.i.d} samplings are represented by arrows. 
To indicate that certain steps in the generative process are replicated, a labelled box or plate is drawn around the relevant parts of the diagram, with the integer label representing  the number of times these steps are to be repeated (thus such graphical models are often also referred to as plate diagrams). 
The graphical mixture model described here is shown in the upper diagram in Figure \ref{graph_models}. 
We can see there that the free parameters of the mixture model given by the theme proportions $\omega_t$ and the multinomial probabilities $\beta_{t,m}$, located outside all plates, have to be defined for the whole event sample in order to initiate the generative process that leads to the measurements $o_{j,i}$ in the inner-most plate.

In collider physics, it is implicit that event samples arise from a statistical mixture of multiple underlying physical scattering process, where  each event is a result of one such particular scattering process. 
Once the corresponding differential cross-sections are binned, the scattering processes can be identified with themes in a multinomial mixture model as described above. 
Traditionally, the weights $\omega$ are computed from first principles using a combination of Quantum Field Theory, Montecarlo event generators tuned to data and experimental knowledge of the detector response. 

More recently, mixture models have been used for semi-supervised classification of event samples where the mixture proportions of the themes are a priori unknown. 
For instance, in the CWoLa framework \cite{Metodiev:2017vrx} a set of event samples $M_1$ and $M_2$ are taken as mixtures of two underlying themes (with different mixing proportions) $S$ and $B$, corresponding to signal and background themes. 
Along this same line, a mixture model was used in Ref.~\cite{Metodiev:2018ftz} with the aim of disentangling the jet substructure distributions (e.g. constituent multiplicity) of quark and gluon jets using mixed event samples. 
The same model and technique was also used in ref.~\cite{Alvarez:2019knh} in an attempt to separate $pp\to t\bar t t\bar t$ from backgrounds in inclusive same-sign dilepton events using jet multiplicity distributions.

There are several drawbacks when using mixture models for (unsupervised) event classification tasks. These come from the assumption that all measurements in an individual event are drawn from one theme. The main (related) issues are the following:
\begin{itemize}
\item Measurements on a single collider event typically receive contributions from many sources, for example in measuring $t\bar{t}$ production it is inevitable that much of the measurements will be of soft QCD radiation rather than the hard decays products of the top quarks.
Mixture models fail to differentiate between different underlying processes in individual events.
\item Mixture models are not well suited for modelling datasets where events generated from different themes share common features.  
Admittedly this is a problem in extracting the themes with the approximate inference techniques, but is a drawback nonetheless, see e.g. ref.~\cite{Blei03latentdirichlet}.  
\end{itemize}  

In general, mixtures are useful representations of the data if the mixing proportions $\omega$ can be computed from first principles or estimated with other means (such as in (semi)supervised ML), but tend to be less suitable if $w$ are a-priori unknown, as in unsupervised ML. Next we discuss mixed-membership models, which address these issues in an efficient way.

\subsubsection{Mixed-membership models}
%%%%%%%%%%%%%%%%%%%%%%%%%%%%%%%%%%%%%%%%%%%%%%%%%%

\noindent 
Mixed-membership models also consist of $T$ themes, however a single event is now generated from a mixture of themes rather than being generated from a single theme, as in the mixture model.
Each event $e_j$ now has its own theme weights $\omega_j=(\omega_{j,1},\ldots,\omega_{j,T})$.
These are now latent variables of the model (not parameters) that are sampled from a prior distribution $p(\omega|\alpha)$, with $\alpha$ being the parameters of the distribution. {This prior is in general defined over the ($T-1$)--dimensional simplex describing the space of all theme weights (i.e. the space of $T$-vectors with positive entries that sum up to one).}  
The generative process for a mixed-membership model goes as follows:

\begin{enumerate}[label=\roman*.]
\itemsep0em
\item Randomly sample a set of $T$ theme proportions from the prior, $\omega_{j}\sim p(\omega|\alpha)$.
\item Randomly sample a theme $t\sim p(t|\omega_j)$.
\item Randomly sample a measurement $o_{j,i}\sim p(o|t,\beta)$.
\item Repeat steps (ii)-(iii) for each measurement in the event.
\item Repeat steps (i)-(iv) for each event in the sample.
\end{enumerate}
We can derive the mixed-membership representation of the joint probability by taking Eq. \eqref{definetti} and assigning
\beq
p(o_{j,i}|\omega)=\sum_{t=1}^Tp(t|\omega_{j})~p(o_{j,i}|t,\beta),
\eeq
where $p(t|\omega_j)=\omega_{j,t}$. Therefore the mixed-membership model for an event is defined by
\beq\label{eq:lda}
p(e_j)=\int_{\Omega} d\omega~ p(\omega|\alpha)\prod_{i=1}^{N_j}\sum_{t=1}^Tp(t|\omega_{j})~p(o_{j,i}|t,\beta)\,,
\eeq
where $\Omega$ is the simplex.
Note the slight change in notation: the latent space variable $\theta$ from Eq. \eqref{definetti} has been replaced with $\omega \leftarrow \theta$ (and $\Omega \leftarrow \Theta$) to keep the notation for mixed-membership models in line with the notation for mixture models. 
The generative process for the mixed-membership model is shown in the lower diagram of Figure~\ref{graph_models}. 
In comparison to the mixture model plate diagram, notice that the theme selection step is now inside the event plate indicating the mixed-membership nature of the model. 
The free parameters of the mixed-membership model are $\alpha$ from the prior and the multinomial probabilities $\beta_{t,m}$ of the themes.

Using mixed-memberships resolves the problem mixture models have when modelling events sharing similar features. 
It is therefore possible to model events that are much more heterogenous. 
It is also clear that the model can now describe events where measurements receive contributions from multiple sources, accommodated now by each event having measurements in a single event sampled from different themes. 

%%%%%%%%%%%%%%%%%%%%%%%%%%%%%%%%%%%%%%%%%%%%%%%%%%
\subsection{Latent Dirichlet Allocation}\label{sec:lda}
%%%%%%%%%%%%%%%%%%%%%%%%%%%%%%%%%%%%%%%%%%%%%%%%%%

\noindent 
In Bayesian probabilistic modelling the inference of model parameters is one of the primary tasks, and will be discussed in detail in Sec. \ref{sec:approximateInference}.
Choosing the prior $p(\omega|\alpha)$ to be the conjugate distribution to the likelihood function makes the parameter inference easier.
For mixed-membership models with a multinomial likelihood, as we have here, the conjugate prior is the Dirichlet distribution $\mathcal{D}(\cdot|\alpha$).
Choosing $p(\omega|\alpha)$ in Eq. \eqref{eq:lda} to be a Dirichlet distribution leads us to Latent Dirichlet Allocation (LDA)~\cite{Pritchard945, Blei03latentdirichlet}.
The Dirichlet distribution defined over the simplex $\Omega$ is a multivariate generalisation of the beta distribution over the unit interval $[0,1]$, reducing to the beta distribution for $T\!=\!2$.
It is in fact a parametric family of distributions, defined by $T$ positive non-zero parameters, $\alpha=\alpha_1,\ldots,\alpha_T$, and has the explicit form 
\beq
\mathcal{D}(\omega|\alpha)=\frac{\Gamma\!\left(\sum_{t=1}^T\alpha_t\right)}{\prod_{t=1}^T\Gamma(\alpha_t)}\,\prod_{t=1}^T\omega_t^{\alpha_t-1}\,,
\eeq
where $\Gamma(x)$ is the gamma function. In LDA, the Dirichlet prior encodes prior information on how we expect the themes to contribute both to individual events and to the whole sample of events. 
It does this by influencing the possible proportions $\omega_j$ selected in the generative process.
For example a particular choice of parameters ($\alpha$) could define a model in which one particular theme contributes much less to individual events than another, or it could define a model in which some events are composed almost exclusively of one theme while others are more equal mixtures of several themes.

\begin{figure}[t!]
\begin{center}
\includegraphics[width=0.8\linewidth]{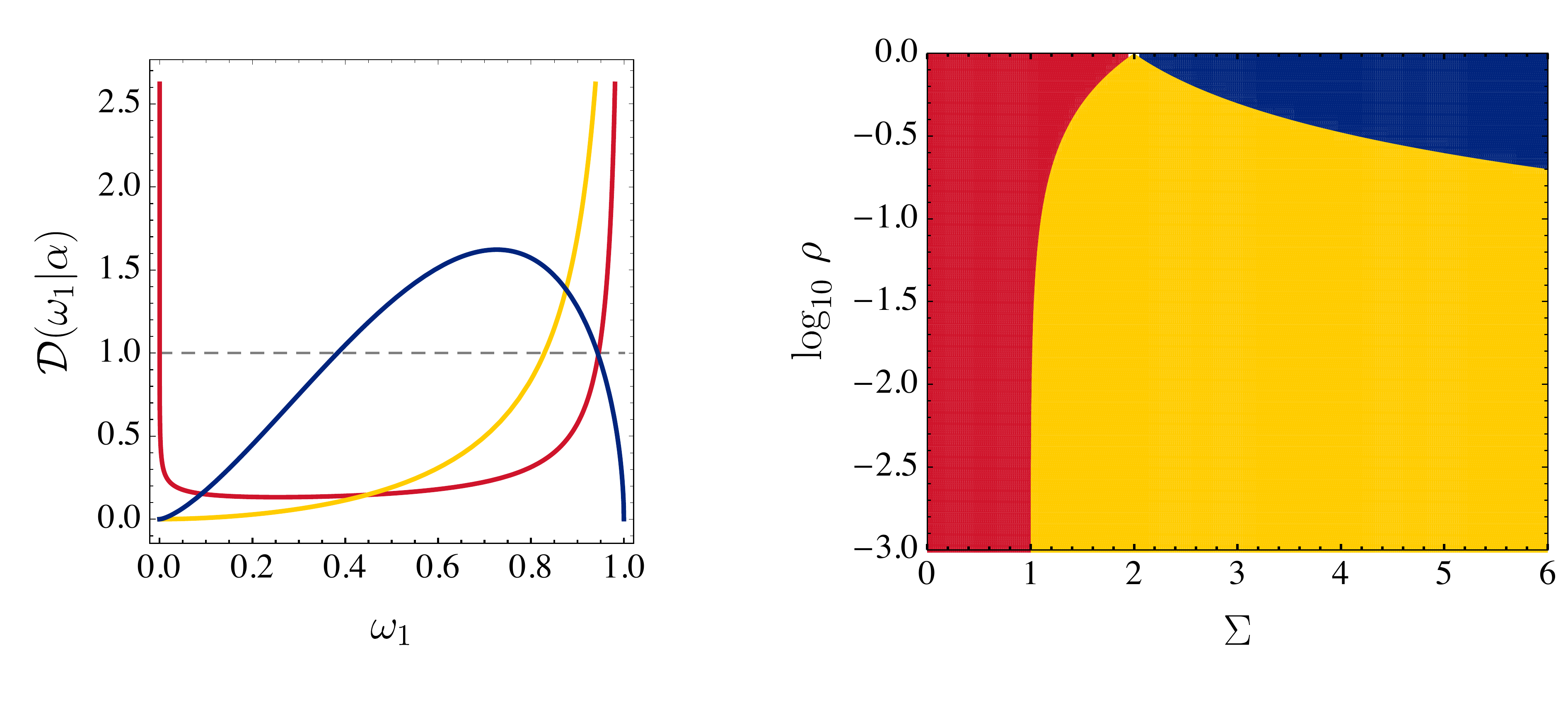}
\vspace{3mm}
\caption{(Left) Three representative Dirichlet priors for $T=2$ over the unit interval drawn in different coloured full lines. The uniform prior $\mathcal D (\omega_1| 1,1)$ is also shown in dashed grey. (Right) Regions with the same prior shapes projected onto the $(\Sigma,\rho)$-plane defined in  eq.~\eqref{new_param}.  The distinct shape inside each colored region is represented by one distribution in the left panel with matching color codes. \label{Beta_distribution}}
\end{center}
\end{figure}

As mentioned in the introduction, we will be concerned solely with scenarios in which only two themes are relevant. 
We will therefore focus on the $T\!=\!2$ case where the Dirichlet prior reduces to a beta distribution. 
For each event we sample a variable $\omega_1$ from the Dirichlet (beta) distribution representing the proportion of the first theme $p(o|1,\beta)$, while the proportion of the second theme $p(o|2,\beta)$ is given by $\omega_{2}=1-\omega_1$. 
In this two-theme scenario, the analytical form of the Dirichlet is given by 
\beq
\mathcal{D}(\omega_1|\alpha_1,\alpha_2)=\frac{\Gamma(\alpha_1+\alpha_2)}{\Gamma(\alpha_1)\,\Gamma(\alpha_2)}\ \omega_1^{\alpha_1-1}(1-\omega_1)^{\alpha_2-1},
\eeq
where for now we drop the $j$ subscript labelling the event.
When inspecting the above distribution family for different values of the $\alpha$ parameters, one identifies several cases that give rise to different types of distribution shapes. 
These different shapes encode different assumptions about underlying event data. 
For instance $\mathcal{D}(\omega_1|1,1)$ corresponds to the flat distribution over the unit interval and would describe events for which the occurrence of either theme in an event is completely random (shown in gray dashed line in Fig.~\ref{Beta_distribution}). 
The other more interesting shapes are the following: 
\begin{enumerate}
\item $\alpha_1\!<\!1,~\alpha_2\!<\!1$: bi-modal distributions (shaded in red in Fig.~\ref{Beta_distribution}) with two maxima at the boundaries of the unit interval ($\omega_1=0$ and $\omega_1=1$). 
Physically, this scenario describes samples for which one group of events have measurements predominantly sampled from the first theme, and another group for which measurements are mostly sampled from the second theme. 
The relative size between each group of events is controlled by the ratio $\alpha_2/\alpha_1$.

\item $\alpha_1\!>\!1,~\alpha_2\!<\!1$:  uni-modal distributions with a maximum located at one boundary of the interval and the distribution tail stretching towards the opposite boundary (shaded in yellow in Fig.~\ref{Beta_distribution}). 
In this case we expect most events to be generated mostly by one predominant theme.

\item $\alpha_1\!>\!1,~\alpha_2\!>\!1$:  uni-modal distributions with one maximum located at $\omega_1=\frac{\alpha_1-1}{\alpha_1+\alpha_2-2}$ and two tails stretching towards both boundaries of the interval (shaded in blue in Fig.~\ref{Beta_distribution}).
In this case we expect the bulk of events to be generated by non-negligible mixtures of both themes, with very few events where just one theme completely dominates.  
However the exact distribution depends strongly on the hierarchy between $\alpha_1$ and $\alpha_2$.
\end{enumerate}
In the following sections we will rely on a useful re-parameterisation of the Dirichlet where we trade the $(\alpha_1,\alpha_2)$ parameters for  $(\Sigma,\rho)$ defined as

\beq\label{new_param}
\Sigma\ \equiv\ \alpha_1+\alpha_2\,,~~~\rho\ \equiv\ \frac{\alpha_2}{\alpha_1}\,.
\eeq

\noindent By convention we have fixed here $\alpha_2\le\alpha_1$, hence $0<\rho\le1$. The $\rho$ parameter controls the asymmetry in the shape of the Dirichlet distribution.
In Fig. \ref{Beta_distribution} (right) we present a visualisation of the the different shapes taken by the Dirichlet distribution, in terms of these $\rho$ and $\Sigma$ parameters.
The smaller $\rho$ is, the more probable events will be composed of measurements drawn from the first theme ($t=1$). 
A way to see this, is by considering the expectation for sampling the themes from the Dirichlet during one measurement sampling. 
One finds
\beq
\mathbb{E}_{\mathcal{D}}\left[\, p_1\omega_1 + p_2(1-\omega_1)\right] \ =\  \int_0^1d\omega_1~ \mathcal{D}(\omega_1|\alpha)\left[\omega_1 p_1+(1-\omega_1)p_2\right]\ =\ \frac{p_1+\rho~\! p_2}{1+\rho}\,,
\eeq
where $p_{t}$ are shorthand for the theme multinomials $p(o|t,\beta_{t,m})$ and $\mathbb{E}_\mathcal{D}[\cdot]$ denotes the expectation with respect to the Dirichlet distribution. To derive this we have used the relation for the mean value $\mu=\mathbb{E}_\mathcal{D}[\omega_1]=\frac{1}{1+\rho}$.
This indicates that in the limit $\rho\to0$ of asymmetric Dirichlet priors, there will be a prevalence of the first theme over the second theme when sampling measurements for an event, while in the limit $\rho\to1$ the priors become symmetric and events will tend on average to have measurements coming from both themes in similar proportions. The parameter $\Sigma$, on the other hand, controls to what degree individual events in the model are described by mixtures of themes for a fixed value of the asymmetry parameter $\rho$, i.e. to what degree the model is a mixed-membership rather than just a mixture model. For large $\Sigma$ we expect that events are generated from mixtures of both themes, whereas for $\Sigma\ll1$ we expect that events are generated from pre-dominantly one theme.
In fact, it is known that the Beta distribution will approach the Bernoulli distribution in the limit of $\Sigma\rightarrow0$ with fixed $\rho$.
In general, a Dirichlet for $T$ themes will approach a $T$-dimensional multinomial distribution in the limit $\sum_{t=1}^T\alpha_t\rightarrow 0$ \cite{betalim1}.
In this limit the Bernoulli probability $p$ is equal to the expectation value of the Dirichlet, $\frac{1}{1+\rho}$.
Therefore in the $\Sigma\ll1$ limit each event is approximately generated by just one theme, and the LDA mixed-membership model tends to the mixture model described previously.
What happens is that when for every event you sample $(\omega_{j,1},\omega_{j,2})$ from the Dirichlet, the only weights that have a non-zero probabilities in the Dirichlet distribution are $(\omega_{j,1}=1,\omega_{j,2}=0)$ and $(\omega_{j,1}=0,\omega_{j,2}=1)$, where the probabilities for selecting each of these from the Dirichlet is $1/(1+\rho)$ and $\rho/(1+\rho)$, respectively.
In this mixture model limit $\rho$ takes on the role of the ratio of theme weights $\omega_1/\omega_2$.
In Fig. \ref{Beta_distribution} we can then identify the boundary at the y-axis as a mixture model with $\omega_1/\omega_2=\rho$.

\begin{figure}[t!]
\begin{center}
\includegraphics[width=0.8\linewidth]{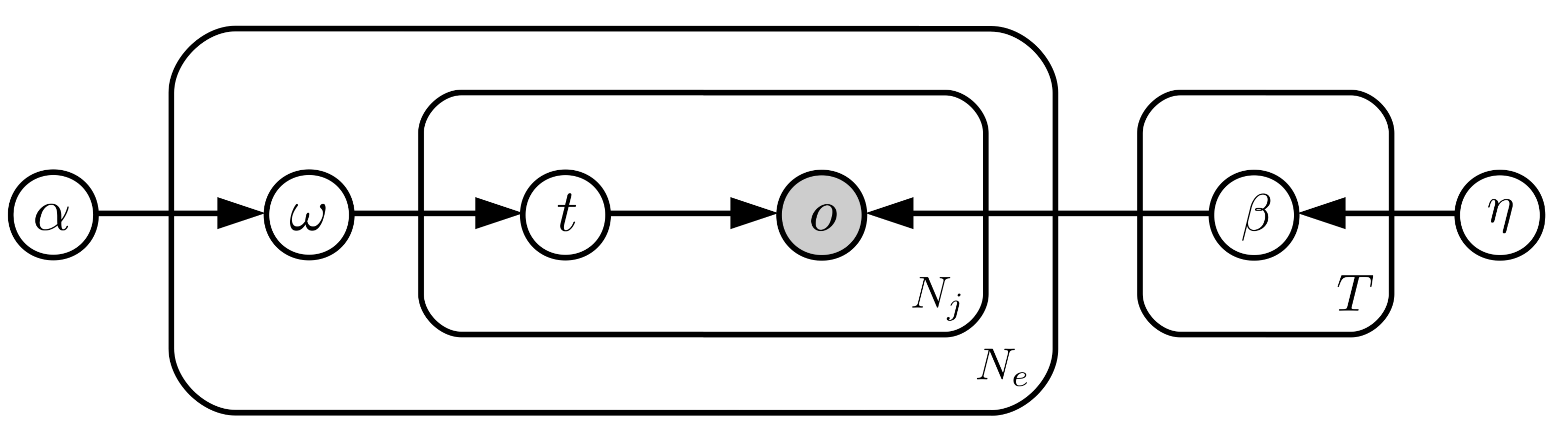}
\caption{The graphical model of smoothed LDA. \label{LDA_graph_model}}
\end{center}
\end{figure}

The event samples we analyse can contain anywhere from $\mathcal{O}(10^3)$ to $\mathcal{O}(10^6)$ events and the number of unique measurements can also be very large.
This means that for parts of the event sample the data will be very sparse, i.e. there will be many $o_{j,i}$ that do not appear often in the sample.
This can lead to issues in the inference procedure, with these rare measurements being assigned zero probability in the themes, which then leads to problems during the classification of events.
This issue can be solved by so-called `smoothing' \cite{Blei03latentdirichlet}.
Smoothing involves placing a $M$-dimensional Dirichlet prior on the variables of the theme probability distributions, such that no measurement can have a zero probability.
The generative process is then augmented as shown in the plate diagram in Fig. \ref{LDA_graph_model}.
We fix each of the $M-1$ parameters of the Dirichlet prior to $1/M$ as default, changing this does not lead to significant changes in the output of the algorithm. Henceforth we will focus on smoothed LDA and refer to it simply as LDA.

%%%%%%%%%%%%%%%%%%%%%%%%%%%%%%%%%%%%%%%%%%%%%%%%%%
\subsection{Approximate inference}
\label{sec:approximateInference}
%%%%%%%%%%%%%%%%%%%%%%%%%%%%%%%%%%%%%%%%%%%%%%%%%%

\noindent  Ultimately, the goal is to estimate the posterior distributions for the variables in the LDA model given the observation of experimental data.
The joint probability over all events $e=(e_1,\ldots,e_{N_e})$ for the LDA model can be written as
\beq
p(e,\beta,\omega,t|\alpha,\eta)\propto \left(\prod_{j=1}^{N_e} p(\omega_j|\alpha) \right) \left(\prod_{t=1}^{T} p(\beta|\eta) \right) \prod_{j=1}^{N_e} \prod_{i=1}^{N_j} p(t_{j,i}|\omega_j)p(o_{j,i}|t_{j,i},\beta),
\eeq
where we have not marginalised over the model variables.
On the left hand side of this equation $\omega$ represents the list of theme weights for all events in the sample, and $t$ represents the list of topic assignments for each $o_{j,i}$ in all events in the sample.
The joint probability is the probability of having generated these events given the LDA model with the themes each being sampled from a Dirichlet parameterised by $\eta$, and the theme weights being sampled per event from a Dirichlet parameterised by $\alpha$.
From this we want to approximate the posterior distribution $p(\beta,\omega,t|e)$.
Bayes theorem states that this posterior should have the form $p(\beta,\omega,t|e)\propto p(e,\beta,\omega,t)/p(e)$.
The term in the numerator is calculable, however the difficulty lies in the normalisation term, the evidence.
This term is an intractable integral and prevents us from straightforwardly obtaining a closed form expression for the posterior distribution.
We can however obtain an approximation to the posterior distribution using approximate inference techniques. Following~\cite{Hoffman:2010:OLL:2997189.2997285, Blei_2017} we choose the Variational Inference (VI) technique.
With VI the log of the evidence is written as
\begin{align}\label{eq:vi}
\ln p(e) &= \int d\xi~ q(\xi)\ln\frac{p(e,\xi)}{q(\xi)}+\int d\xi~q(\xi)\ln\frac{q(\xi)}{p(\xi|e)} \nonumber \\
&=\mathcal{L} + \text{KL}(q(\xi)\big|\big| p(\xi|e))\,,
\end{align}
where $\rm KL$ stands for the Kullback-Leibler divergence~\cite{Kullback} and we use $\xi$ as a shorthand for the model variables, ($\beta,\omega,t$).
The function $q(\xi)$ has been introduced here as an approximation to the posterior distribution,
\beq
p(\beta,\omega,t|e)\simeq q(\beta,\omega,t)\equiv q(\beta)q(\omega)q(t)\,,
\eeq
where $q(\beta,\omega,t)$ is assumed to factorise in each variable, reflecting how these are grouped in LDA.
The goal of VI is to approximate this $q(\xi)$ function.
On the right-hand-side of Eq. \eqref{eq:vi} we have two terms: the Evidence-Lower-BOund (ELBO) $\mathcal{L}$, and the KL divergence between the posterior and its approximation.
The KL divergence is always greater than zero, and is equal to zero only when $q(\beta,\theta,t)=p(\beta,\theta,t|e)$.
The term $\mathcal{L}$ is then a lower bound on the evidence, hence calling it the ELBO.
We cannot compute the KL divergence because we cannot compute the posterior, however  the joint likelihood and therefore  the ELBO term can be computed.
The goal is then to maximise the ELBO with respect to $q(\beta,\omega,t)$. 
Because the evidence term on the right-hand-side is completely independent of $q(\beta,\omega,t)$, maximising the ELBO is equivalent to minimising the KL divergence between $q(\beta,\omega,t)$ and the posterior, thus finding a $q(\beta,\omega,t)$ which is a good approximation to the posterior.
VI gives us a prescription for doing this in mixed-membership models like LDA.

The LDA model belongs to the conjugate exponential family of models. For these, one can show that the terms in the posterior approximation must have the following form:
\begin{align}
q(t_{j,i})=&\,\text{Multinomial}(\phi_{j,i}),~~(j=1,\ldots,N_e),~(i=1,\ldots,N_j)\,,\nonumber \\
q(\omega_j)=&\,\text{Dirichlet}(\alpha_{j,t}),~~(j=1,\ldots,N_e),~(t=1,\ldots,T)\,, \nonumber \\
q(\beta_t)=&\,\text{Dirichlet}(\gamma_{t,m}),~~(t=1,\ldots,T),~(m=1,\ldots,M).
\end{align}
So to optimise $q(\beta,\omega,t)$ we need to maximise the ELBO with respect to the parameters $\phi_{j,i},\alpha_{j,t},\gamma_{t,m}$.
Note that there are $T(n_e\!+\!M)\!+\!N_J$ parameters here, where $N_J$ is the total number of measurements in all events in the sample.
Due to the specific structure of LDA, i.e.~the conditional dependencies and the use of conjugate priors, closed form expressions of the parameters that maximise the ELBO can be written in terms of each other (see below).
The VI algorithm then dictates how to update the parameters iteratively such that it converges to a maximum of the ELBO function.

Due to the large number of events from which we infer the parameters of the approximate posterior, the basic VI algorithm is inefficient.
To implement this in a way which scales well to large datasets we employ an extension of this algorithm called Stochastic Variational Inference (SVI).
This technique uses results from stochastic optimisation methods to speed up the inference by inferring from smaller randomly sampled subsets of the data on each update.
These are called chunks of data, and their size is determined by the chunk size $n_c$.
The algorithm will run for a total number of passes through the dataset, defined by $n_p$.
We denote the total number of chunks of data processed by $N$.
The algorithm is thoroughly defined as follows:\\
\begin{itemize}
\item{\bf Inputs}\newline Event data, and the approximate posteriors $q(t_{j,i})=\text{Multinomial}(\phi_{j,i})$, $q(\omega_j)=\text{Dirichlet}(\alpha_{j,t})$, $q(\beta_t)=\text{Dirichlet}(\gamma_{t,m})$.
\item{\bf Outputs} \newline Posterior distributions for $\alpha_{j,t}$, $\phi_{j,i}$, and $\gamma_{t,m}$.
\item{\bf Procedure}
\begin{enumerate}
\item Initialise $\gamma_{t,m}^{(n=0)}$.
\item For chunks $n=1,\ldots,N$, do:
\begin{enumerate}
\item Initialise $\alpha_{j,t}^{(l=0)}$, $\phi_{j,i}^{(l=0)}$\,.
\item For iterations $l=1,\ldots,L$ do:
\begin{enumerate}
\item Update $q(\phi_{j,i})$ by iterating through $j$ and $i$ and setting
\beq\label{eq:phiupdate}
\phi^{(l)}_{j,i}(t)=\frac{ e^{ \psi\left(\gamma_{t,o_{j,i}}^{(n-1)}\right) - \psi\left(\sum_{m=1}^M\gamma^{(n-1)}_{t,m}\right)) + \psi\left(\alpha_{j,t}^{(l-1)}\right) - \psi\left(\sum_{p=1}^T\alpha_{j,p}^{(l-1)}\right) } }{ \sum_{s=1}^T e^{ \psi\left(\gamma_{s,o_{j,i}}^{(n-1)}\right) - \psi\left(\sum_{m=1}^M\gamma^{(n-1)}_{s,m}\right)) + \psi\left(\alpha_{j,s}^{(l-1)}\right) - \psi\left(\sum_{p=1}^K\alpha_{j,p}^{(l-1)}\right) }}\,.
\eeq
\item Update $q(\omega_j)$ by iterating through $t$ and setting:
\beq\label{eq:alphaupdate}
\alpha_{j,t}^{(l)}=\alpha_t+\sum_{i=1}^{N_j}\phi_{j,i}^{(l)}(t)\,.
\eeq
\item Check for convergence: if the change in $\alpha$ is less than the threshold parameter $\alpha_{\text{thresh}}$, end loop.
\item Set $\phi_{j,i}^{(n)}=\phi_{j,i}^{(L)}$ and $\alpha_{j,t}^{(n)}=\alpha_{j,t}^{(L)}$.
\end{enumerate}
\item Update the themes.
\begin{enumerate}
\item Update $q(\beta_t)$ by iterating through $t$ and $m$ and setting:
\beq\label{eq:gammaupdate}
\gamma_{t,m}^{(n)}=(1-\delta_n)\gamma_{t,m}^{(n-1)} + \delta_n\left(\eta+\sum_{j=1}^{N_e}\sum_{i=1}^{N_j}\phi_{j,i}^{(n)}(t)\mathbb{I}(o_{j,i}=m)\right)\,.
\eeq
\end{enumerate}
\item Evaluate the normalised (per-$o_{j,i}$) ELBO for this chunk of data from the dataset, $\mathcal{L}_n$.
This can be used to check for convergence.\end{enumerate}
\end{enumerate}
\item{\bf Return}\newline  $\alpha_{j,t}^{(N)}$, $\phi_{j,i}^{(N)}$, and $\gamma_{t,m}^{(N)}$.\\
\end{itemize}
The algorithm makes use of the hierarchical structure of the model, with local variables ($\omega,t$) being optimised until convergence before an update on the global variables (themes, $\beta$) is performed.
While optimising the local variables the algorithm uses the $(l-1)^{\text{th}}$ approximation of $\alpha_{j,t}$ and the $(n-1)^{\text{th}}$ approximation of $\gamma_{t,m}$ to calculate the $l^{\text{th}}$ approximation of $\phi_{j,i}$, before using the $l^{\text{th}}$ approximation of $\phi_{j,i}$ to calculate the $l^{\text{th}}$ approximation to $\alpha_{j,t}$. Once $L$ updates of this sort have been done, or until convergence has been met according to $\alpha_{\text{thresh}}$, the themes are updated using the local variables obtained at the end of the inner loop.

A few points of note: (i) the $\psi(\cdot)$ in Eq. \eqref{eq:phiupdate} arises from the expectation of the natural logarithm of the Dirichlet distribution, (ii) the $\eta$ in Eq. \eqref{eq:gammaupdate} is from the prior on the theme distributions, (iii) the $\mathbb{I}(\cdot)$ in Eq. \eqref{eq:gammaupdate} is an indicator function, which is equal to $1$ when the equality in the brackets is true, and equals zero when it is not. 

It is also important to note the key role played by the latent variable $\phi_{j,i}$, which encodes information on which theme each measurement in each event was sampled from. This variable captures the co-occurrences between different measurements in the event sample. For example, if some measurement $m'$ co-occurs with another measurement $m''$ in many events, this information is stored by the $\phi_{j,i}$ variable and through iterative updates these two measurements are more likely to end up with large weights in the same theme distribution.
{It is through the presence of co-occurring measurements in the data that this algorithm is able to disentangle different underlying physical processes occurring in the events.
Without these co-occurrences, or a method to utilise them, the best an unsupervised algorithm can do in identifying rare events is to search for outliers in the data.
Thus searching for these co-occurrences is essential in extracting a generative description of the events.
We pay particular attention to this in deciding upon a data representation (Sec.~\ref{sec:fc}) for our benchmark studies.}

In this work we have used the implementation of the SVI procedure as described above within \texttt{gensim}~\cite{rehurek_lrec}, a software package for performing unsupervised semantic modelling of plain text. The parameters of the SVI algorithm are the chunk size $n_c$, the number of iterations $L$, the alpha threshold $\alpha_{\text{thresh}}$, and the number of passes $n_P$. 
On the other hand, the learning rate $\delta_n$ is not constant in \texttt{gensim} but follows
\beq
\delta_n=\frac{1}{\left(\tau_0+n\right)^\kappa}\,,
\label{eq:learn}
\eeq
where $\tau_0$ is the offset and $\kappa$ is the decay parameter.
This stochastic inference procedure is proven to converge to a local minimum if $\sum_{n=1}^\infty \delta_n=\infty$ and $\sum_{n=1}^\infty \delta_n^2<\infty$, which is guaranteed for $\kappa \in (\tfrac{1}{2},1]$.
The convergence of the whole algorithm can be assessed using the ELBO, or equivalently, the perplexity defined as
\beq\label{eq:perp}
\mathcal{P}_n=2^{-\mathcal{L}_n}.
\eeq
A lower perplexity means that the ELBO is larger and thus the KL-divergence between the posterior and approximate posterior is smaller.
In Sec. \ref{sec:appsys} we study how the choices of the offset and the chunk size parameters of the algorithm affect the convergence and performance of the models as well as their final perplexity.

With the posterior distributions at hand, we would typically like to infer the theme distributions and the theme weights of individual events.
Ideally, we would maximise the posterior distributions with respect to the variational parameters to obtain best estimates of the theme parameters and mixing weights for the LDA model, however this is computationally difficult \cite{tminka1}.
A good approximation for the theme parameters and mixing weights can instead be obtained by simply taking the expectation values, 
\begin{align}
\hat{\beta}_{t,m}=&\,\mathbb{E}_q[\beta_{t,m}]=\frac{\gamma_{t,m}}{\sum_{m=1}^M\gamma_{t,m}}\,, \\
\hat{\omega}_{j,t}=&\,\mathbb{E}_q[\omega_{j,t}]=\frac{\alpha_{j,t}}{\sum_{t=1}^T\alpha_{j,t}}\,.
\end{align}

%%%%%%%%%%%%%%%%%%%%%%%%%%%%%%%%%%%%%%%%%%%%%%%%%%
\subsection{Event classification}
%%%%%%%%%%%%%%%%%%%%%%%%%%%%%%%%%%%%%%%%%%%%%%%%%%

\noindent Once we have the posterior approximation and the estimates of the theme distributions $(\hat{\beta}_{t,m})$ and the theme weights for each event $(\hat{\omega}_{j,t})$, we want to be able to use this information to cluster events into one of two clusters, $\mathcal{C}_1$ or $\mathcal{C}_2$.
The mixed-membership model assumes that each event is already a mixture of two types of underlying themes, so we could simply cluster the events by placing cuts on $\hat{\omega}_{j,1}$ for each event:
\begin{align}
\hat{\omega}_{j,1}&> c~\Rightarrow~e_j\in \mathcal{C}_1\,, \nonumber\\
\hat{\omega}_{j,1}&\leq c~\Rightarrow~e_j\in \mathcal{C}_2\,.
\end{align}
Classifying events in this way does yield good classification performance, as demonstrated in our earlier work \cite{Dillon:2019cqt}. However, we have also found that using instead the likelihood-ratio classifier yields a more robust performance over a larger region of model prior space.
The likelihood ratio classifier is constructed using just the themes as extracted from the data, and not the theme weights.
The likelihood ratio can be written as
\begin{align}\label{eq:lr}
L(e_j)=L(o_{j,i})&=\prod_{i=1}^{n_j}\frac{p(o_{j,i}|\beta_{2})}{p(o_{j,i}|\beta_{1})} \nonumber \\
&=\prod_{m=1}^{M}\left(\frac{\beta_{2,m}}{\beta_{1,m}}\right)^{\mathbb{I}(m=o_{j,i})}\,,
\end{align}
where $\mathbb{I}(\cdot)$ is again the indicator function, equal to $1$ when the expression in brackets is true and equal to $0$ when it is not.
With the likelihood ratio we also need to perform a cut in order to cluster the events,
\begin{align}
L(e_j)&\leq c~\Rightarrow~e_j\in \mathcal{C}_1\,, \nonumber\\
L(e_j)&> c~\Rightarrow~e_j\in \mathcal{C}_2\,.
\end{align}

\subsubsection{Evaluating the performance of a model}

\noindent We can evaluate how well a particular classification technique performs using Monte-Carlo generated data, for which we know the truth labels.
Suppose we generate two samples of events, sample 1 and sample 2, and we produce a mixed sample of events from both pure samples.
We can train an LDA model with $T=2$ on this mixed sample to extract $2$ theme distributions that describe the data.
We can then either use the extracted theme weights, or the likelihood ratio to cluster the events in either $\mathcal{C}_1$ or $\mathcal{C}_2$\,.
Suppose the goal is to cluster events from sample 1 into $\mathcal{C}_1$, and events from sample 2 into $\mathcal{C}_2$. We can test how well the algorithm performs using the truth labelled data.
We can compute the fraction of events from sample 2 correctly assigned to $\mathcal{C}_2$ as a function of the cut, $\varepsilon_2(c)$.
And analogously we can compute the fraction of events from sample 1 incorrectly assigned to $\mathcal{C}_2$ as a function of the cut, $\bar{\varepsilon}_1(c)$.
The Receiver-Operating-Characteristic (ROC) curve is then defined as the curve tracing the true positive rate as a function of the false positive rate, i.e. $\varepsilon_2\left(\bar{\varepsilon}_1\right)$.
Two measures we use to evaluate the performance of the LDA models we have trained are
\begin{enumerate}
\item Area Under Curve (AUC):  the integrated area under the ROC curve.
\item Inverse mistag at fixed efficiency: $\bar{\varepsilon}_1^{-1}\left(\varepsilon_2=0.5\right)$.
\end{enumerate}
The AUC is a useful statistic when we are interested in the overall general performance of the classifier.
However when the experimental analysis is focused on identifying rare signals in a sample of events, the AUC statistic is not always the most relevant indicator of performance.
What is required is a statistic which demonstrates a strong rejection of background events coinciding with the acceptance of a moderately large number of signal events.
This is captured by the inverse mistag at fixed efficiency.

The goal is to determine the LDA priors and VI parameters that lead to the best performing models -- the ones best characterising and differentiating the pure samples and at the end offering the best classification performance.  
As we demonstrate in Sec.~\ref{sec:ldaresults} the classification performance is indeed correlated with the perplexity of the model, Eq. \eqref{eq:perp}, when calculated using all the data in the event sample.
A decrease in perplexity corresponds to an increase in the ELBO and therefore a decrease in the KL-divergence between the approximate and true posteriors.
Hence the better we approximate the posterior, the better we expect the classifier to perform.

%%%%%%%%%%%%%%%%%%%%%%%%%%%%%%%%%%%%%%%%%%%%%%%%%%
\section{Learning latent jet substructure}
\label{sec:latentjetsubstructure}
%
%%%%%%%%%%%%%%%%%%%%%%%%%%%%%%%%%%%%%%%%%%%%%%%%%%

\noindent So far we have introduced probabilistic generative generative models as a tool for analysing experimental data, in particular for extracting rare signals in a dataset.
As an example of how this works in practice, we apply LDA to the analysis of di-jet events.
In this section we explain how to represent di-jet events in terms of a sequence of exchangeable measurements $o_{j,i}$, and discuss how the mixed-membership model is well suited for describing di-jet events, and finally we discuss our choice of $o_{j,i}$ representation and basis.

%%%%%%%%%%%%%%%%%%%%%%%%%%%%%%%%%%%%%%%%%%%%%%%%%%
\subsection{Jet de-clustering and substructure observables}\label{sec:jdso}
%%%%%%%%%%%%%%%%%%%%%%%%%%%%%%%%%%%%%%%%%%%%%%%%%%

\noindent When coloured particles are produced at high energy colliders the subsequent QCD showering, fragmentation, and hadronization results in many hadrons in the final state.
If the transverse momentum of the initial particle is large enough, all of these final-state hadrons will be registered by the detector within a single localised region in $(\eta,\phi)$.
These clusters of hadrons are referred to as jets, and there have been many different clustering techniques developed to define jets based on the four-momenta of the constituent hadrons.
Of these different techniques, the sequential recombination schemes \cite{Catani:1993hr, Ellis:1993tq, Cacciari:2008gp, Dokshitzer:1997in, Wobisch:1998wt} have become the standard algorithms for jet clustering. 
When applied to data collected for a single collider event, the algorithm can reduce the complexity of the data to a handful of jets, each representing a final state of some high-energy parton produced in the hard collision. 
In order to arrive at a single clustered jet from hundreds of hadrons, the sequential recombination scheme goes through a set of pairwise intermediate clusterings in which the four-momenta of two subjets are combined to form a larger subjet.
De-clustering the jet and analysing these individual splittings can provide crucial information into the physical processes taking place during the event.
For example, if the initial particle is a top quark with a large transverse momentum, the resulting jet will contain splittings that describe the decay of the top quark into a bottom quark and a $W$ boson, and splittings describing the decay of the $W$ boson. These features are readily exploited by existing traditional top taggers, see e.g.~\cite{Plehn:2010st}.

To analyse di-jet events with the probabilistic generative models outlined in this paper, we extract measurements from each of these splittings in the (de)clustering procedure.
At each splitting we construct a number of observables using the four-momenta of the subjet being de-clustered ($j_0$), and the two subjets resulting from the de-clustering ($j_1$ and $j_2$).
The process for doing this is straight-forward but we must decide on a fixed set of observables to use throughout, this will be discussed in detail in Sec. \ref{sec:sbm}.
Once we have collected a set of measurements at a splitting, we must then bin their values, e.g. according to the detector resolution, but more importantly according to what the algorithm can realistically handle.
The relationship between the size of the observable bins and the algorithm performance is discussed in detail in Sec. \ref{sec:fc}.
One of these binned lists of observables is what we refer to as a measurement $o_{j,i}$ in the probabilistic model.
Because of the binning there will be a finite (although possibly very large) number of values that each $o_{j,i}$ can take.
In addition to the kinematical observables at each splitting we add one more categorical observable, that is a label identifying which jet the splitting belongs to.
With these methods we are describing the whole event rather than a single jet, so the information to which jet a splitting belongs is important to properly characterise the whole event.
Of course, including measurements from all splittings in the jet clustering history is not necessary, and would hinder the VI algorithm in extracting themes relevant for describing a potential signal.
Thus we need to impose cuts such that most of the splittings that are irrelevant to uncovering rare signal events are removed, for example a simple cut on subjet masses removing splittings of subjets with $m_{0}<m_{\text{cut}}$ could remove many of the soft emissions occurring near the end of the QCD showering process.
The whole process, starting from the raw event data, can be described as follows:
\begin{enumerate}
\itemsep0pt
\item Cluster the event with a large jet radius, and keep only the two hardest jets.
\item De-cluster each jet, extracting a list of measurements at each splitting.
\item Bin the measurements from each splitting appropriately, and assign a label identifying which jet the splitting belongs to. 
\item Apply kinematical cuts on the splittings.
\end{enumerate}
An event is then described by an ordered sequence of $o_{j,i}$ each representing a splitting, where each $o_{j,i}$ consists of a list of binned measurements and a label identifying which jet the splitting occurred in.
We would like to point out that this method, and the model in general, does not rely on any specific clustering scheme.
Any set of measurements which describe substructure kinematics of the jets could be used.

\begin{figure}[t!]
\begin{center}
\includegraphics[width=0.9\linewidth]{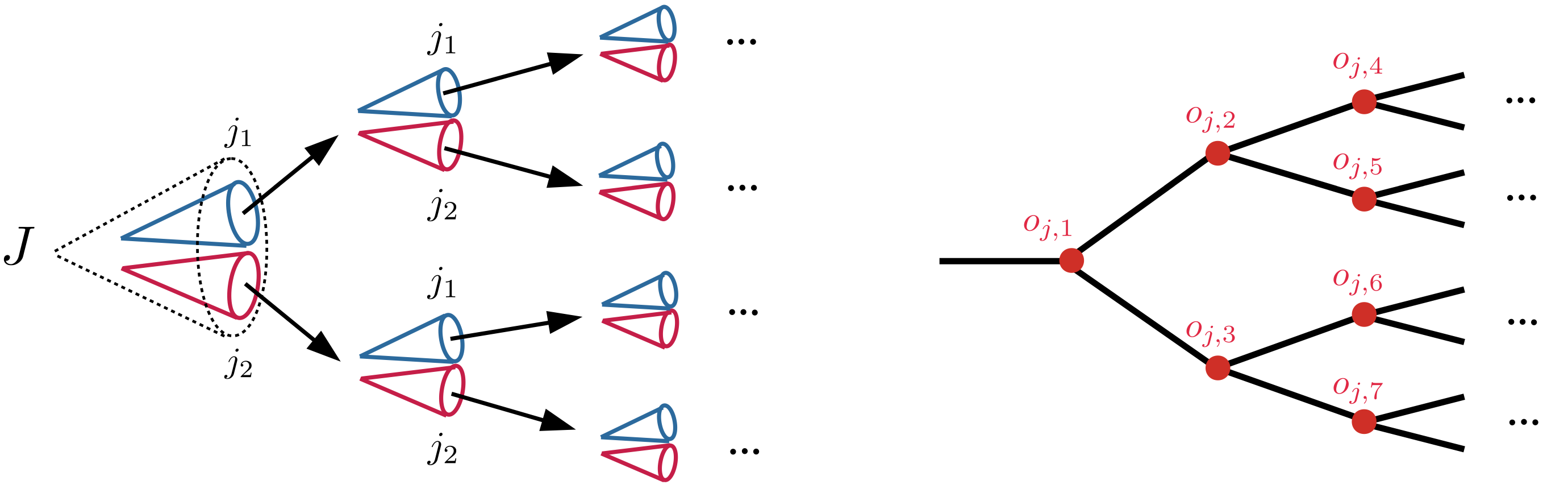}
\caption{ In the left plot we show schematically how the sequential unclustering algorithm proceeds, with the whole jet $J$ being repeatedly separated into two subjets ($j_1$ and $j_2$ with $m_{j_1}>m_{j_2}$).
In the right plot we then show how the feature representation of the data maps onto this unclustering, with each $o_{j,i}$ being mapped to a node in the unclustering tree.
Note that the ordering of these $o_{j,i}$ terms with a single jet does not matter.
 \label{cluster_tree}}
\end{center}
\end{figure}

%%%%%%%%%%%%%%%%%%%%%%%%%%%%%%%%%%%%%%%%%%%%%%%%%%
\subsection{Probabilistic models of jet substructure}
%%%%%%%%%%%%%%%%%%%%%%%%%%%%%%%%%%%%%%%%%%%%%%%%%%

\noindent At the core of the probabilistic models discussed in Sec. \ref{sec:probmodelling} is De Finetti's theorem.
 Under the assumption that the measurements $o_{j,i}$ used to describe the events are exchangeable, this theorem allows us 
to derive, based on additional modelling assumptions, the different latent structures in mixture models and mixed-membership models.
Constructing the $o_{j,i}$ variables for jet substructure as described in the previous section is in line with the exchangeability assumption.
Sequential jet clustering algorithms do impose an ordering on the splittings due to the pairwise nature of the algorithms and the procedure through which the next subjets to be clustered are selected.
However it is the kinematical properties of the splittings that cary most of the interesting physical information, not the order in which they occur, as shown e.g. in~\cite{Dreyer:2018nbf}.

We see then that the latent themes in both the mixture and mixed-membership models for di-jet events are probability distributions over the space of possible splittings (de-clusterings) that can occur within the two leading jets.
The generative processes for the mixture model and mixed-membership (LDA) model are of course different.
In a mixture model a theme would ideally be associated to the specific (hard) partons produced in the collision.
Each splitting in an event described by a mixture model is sampled from just one theme, therefore this theme must represent all of the physical processes occurring within the jets produced within that event.
In a mixed-membership model however, different themes can be associated to different physical processes occurring within the jets of a single event.
Each event in a mixed-membership model is composed of a mixture of themes, just as there are mixtures of different physical processes occurring within each event.
The theme proportions for each event are selected individually from a prior distribution, whose parameters are important in the modelling.
Measurements in each event are `generated' by first sampling theme proportions from the prior, then for each splitting $o_{j,i}$ a theme is drawn from the theme proportions, and a splitting is sampled from that theme.

As an example consider modeling a mixed sample of events consisting of QCD di-jet events ($pp\rightarrow jj$), and top quark pair-production events ($pp\rightarrow t\bar{t}\rightarrow (W^+b)(W^-\bar{b})\rightarrow jj$) where the top quarks are boosted enough such that the decay products of a single top are clustered into a single jet.
The splittings within a QCD jet will be predominantly soft with the number of splittings at higher $k_T$ being monotonically suppressed.
For the top jets the decay chain also involves many coloured particles (the top, the bottom, the decay products of the $W$ boson), therefore there will be many, predominantly soft, gluon emissions occurring within the top jets as well.
However there will always also be a few hard splittings corresponding to the decay of the top quark to the bottom quark and $W$ boson, and the decay of the $W$ boson to light quarks.
Using a two-theme mixture to model this event sample would ideally lead to one theme describing all the splittings within QCD jets, and one theme describing both the hard (decay) splittings and soft (QCD) splittings within the top jets.
With a mixed-membership model on the other hand, the soft splittings occurring within both the QCD and top jets can be modeled by one theme, with the other theme describing just the hard splittings related to the decay dynamics inside top jets. 
This seems like a natural setting in which to search for rare new physics signals in di-jets at high-energy colliders.

%%%%%%%%%%%%%%%%%%%%%%%%%%%%%%%%%%%%%%%%%%%%%%%%%%
\subsection{Choosing a data representation for the jet substructure}
\label{sec:jetreps}
%%%%%%%%%%%%%%%%%%%%%%%%%%%%%%%%%%%%%%%%%%%%%%%%%%

\noindent The discussion so far has not been specific to which observables are to be measured at each $j_0 \to j_1 j_2$ splitting in the jets.
In this subsection we will discuss and justify two bases of observables, with each basis using a different cut to determine which splittings are included in the analysis.
Note that in the end we will only use a subset the observables from each basis, as explained in more detail in Sec. \ref{sec:sbm}.

\noindent The first choice is what we refer to as the mass basis, see e.g.~\cite{Plehn:2009rk, Plehn:2010st}:
\begin{itemize}
\itemsep0pt
\item[] {\it mass-basis}: $\{m_{0},\frac{m_{1}}{m_{0}},\frac{m_{2}}{m_{1}},\frac{k_T}{m_0},\cos\theta\}$~~\text{where $m_{1}>m_{2}$}.
\end{itemize}
These are the mass of the (mother) (sub)jet being de-clustered, the mother/daughter subjet mass drop, the daughter subjets' mass ratio, the $k_T$ distance between the daughter subjets defined in the usual way as $k_T=p_{T,2}\Delta$, where $\Delta^2=(y_2-y_1)^2+(\phi_2-\phi_1)^2$, and the helicity angle between the mother (sub)jet and the daughter subjets as defined e.g. in~\cite{Chwalek:2007pc,Kaplan:2008ie}.
In this basis we only include splittings from the jets in which the subjet being de-clustered has a mass $m_{0}>30$ GeV.

\noindent The second choice is what we refer to as the Lund basis \cite{Dreyer:2018nbf}:
\begin{itemize}
\itemsep0pt
\item[] {\it Lund-basis}: $\{m_{0}, \log R/\Delta, \log k_T, \log R/\kappa, z, \psi\}$,
\end{itemize}
where $R$ is the jet radius, $z=p_{T,2}/(p_{T,1}+p_{T,2})$, $\kappa=z\Delta$, $\psi=\tan^{-1}(y_2-y_1)/(\phi_2-\phi_1)$, and $p_{T,1}>p_{T,2}$.
In this basis we only include splittings from the jets which lie on the primary Lund plane.
The primary Lund plane is defined as the path through the clustering history, starting from the clustered jet, and continually moving through the pairwise splittings to the subjet with the largest $p_T$ until the end of the clustering history. One advantage of the primary Lund plane compared to the mass-basis is that it offers a clearer interpretation in terms of hard vs. soft (i.e. perturbative vs. non-perturbative) splittings, see Ref.~\cite{Dreyer:2018nbf} for details.

We emphasise that these two bases do not just differ in the observables (in fact both bases include the subjet mass $m_{0}$ and (log of) $k_T$), but the different cuts make a considerable difference in the splittings which are used for the description of the jets.
In Sec.~\ref{sec:fc} we explore how some features of the dataset change as we vary the binning used for these observables. Here we only specify the default bin sizes: for the mass-basis observables we bin the measurements in intervals of $\{10\,\text{GeV},0.05,0.05,0.05,0.1\}$\,, while for the Lund basis we use $\{10\,\text{GeV},0.2,0.2,0.05,0.2,0.2\}$\,. 

The last thing to discuss in terms of the data representation are the jet labels.
In Sec. \ref{sec:jdso} we discussed the importance of including jet labels to differentiate between splittings occurring in the two jets, however we did not specify how these jets should be labelled.
Naively, because we select the jets according to $p_T$, one might choose to also label the jets in the same way with $J_1$ being the jet leading in $p_T$ and $J_2$ being the jet subleading in $p_T$, i.e. $p_{T,J_1}>p_{T,J_2}$.
However this is not suitable in practice.
In the top quark pair production example discussed in previous subsection the ordering of the jet labels is not so important, since both jets in the event are top jets and have the same decay structure.
However not all of the signals we may imagine will be so simple.
In many cases, including the new physics example studied in this paper, the two jets in the final state will have been seeded by two different particles of different mass and thus they will both have distinctly different decay dynamics.
Being able to differentiate between these different structures is not just important for classification, but is also important for a physical interpretation of the themes learned through the VI algorithm.
Therefore in the case where the signal events contain two different jets, we would like to be able to associate the $(J_1,J_2)$ labels with splittings from one jet or the other, consistently across the whole sample.
This will not happen if we label the jets by their $p_T$, instead the best way to do this is by labelling the jets according to their jet mass $m_J$, such that $m_{1}>m_{2}$.

%%%%%%%%%%%%%%%%%%%%%%%%%%%%%%%%%%%%%%%%%%%%%%%%%%
%
\section{Set-up and benchmarks}\label{sec:sbm}
%
%%%%%%%%%%%%%%%%%%%%%%%%%%%%%%%%%%%%%%%%%%%%%%%%%%

%%%%%%%%%%%%%%%%%%%%%%%%%%%%%%%%%%%%%%%%%%%%%%%%%%
\subsection{Algorithm set-up}
%%%%%%%%%%%%%%%%%%%%%%%%%%%%%%%%%%%%%%%%%%%%%%%%%%

\noindent There are a number of parameters that determine how the VI algorithm is implemented, these have been discussed in Sec. \ref{sec:approximateInference}. In our benchmark examples we use the following choices, which produce robust results across a wide range of scenarios: passes $n_p= 200$, chunk size $n_c=  10^4$, iterations $L= 100$, offset $\tau_0= 1000$, $\alpha_{\text{thresh}}= 10^{-8}$, decay $\kappa= 0.5$.
These choices are justified in Sec. \ref{sec:appsys} where we discuss in particular how changing the chunk size and the offset affects the convergence of the algorithm, and the performance of the classifier.

%%%%%%%%%%%%%%%%%%%%%%%%%%%%%%%%%%%%%%%%%%%%%%%%%%
\subsection{Benchmark di-jet events}
%%%%%%%%%%%%%%%%%%%%%%%%%%%%%%%%%%%%%%%%%%%%%%%%%%

\noindent We perform our analysis using two benchmark scenarios, (i) boosted top quark pair-production $pp\to t\bar{t}\to b\bar{b}W^+W^-$, and (ii) a hypothetical $3$ TeV vector $W'$ plus a $400$ GeV scalar $\phi$ model, with the dominant production and decay chain $p p \to W' \to W (\phi \to W W)$. 
Since the choices of observables here focus only on the jet substructure, we consider only the hadronic final states of the $W$ bosons in both cases. Consequently, the main background process in both scenarios is the QCD di-jet production.
All event samples were generated using \texttt{aMC@NLO}\citep{Alwall:2014hca} interfaced with \texttt{Pythia 8.2}\cite{Sjostrand:2007gs} for showering and hadronization, and \texttt{FastJet 3.4.1}\cite{Cacciari:2011ma} for jet clustering.
The events were generated at a collision energy of $13$ TeV and the jets were clustered using the CA algorithm\cite{Dokshitzer:1997in, Wobisch:1998wt} with $R=1.5$. No jet grooming was performed. Finally, {for $t\bar{t}$ ($W'$),} jets with $p_T<300$ {($400$)} GeV were discarded.  The detector effects were not simulated, however we checked that the effects of subcluster energy smearing consistent with the \texttt{Delphes 3.4.1}\cite{deFavereau:2013fsa} simulation of the ATLAS detector had no significant effect on the results.

\subsubsection{Boosted top quark pair-production}\label{sec:bm1}
%%%%%%%%%%%%%%%%%%%%%%%%%%%%%%%%%%%%%%%%%%%%%%%%%%

\noindent In the recent years the $pp\rightarrow t\bar{t}\rightarrow b\bar{b}W^+W^-$ process has become a standard benchmark for supervised machine learning applications to particle physics~\cite{Kasieczka:2019dbj}.
Despite there being no need for an unsupervised top tagging algorithm, we find that this is a nice example to demonstrate the power of these techniques as applied to a physical process that is already well measured and understood.

\begin{figure}[t]
  \centerline{\includegraphics[scale=0.45]{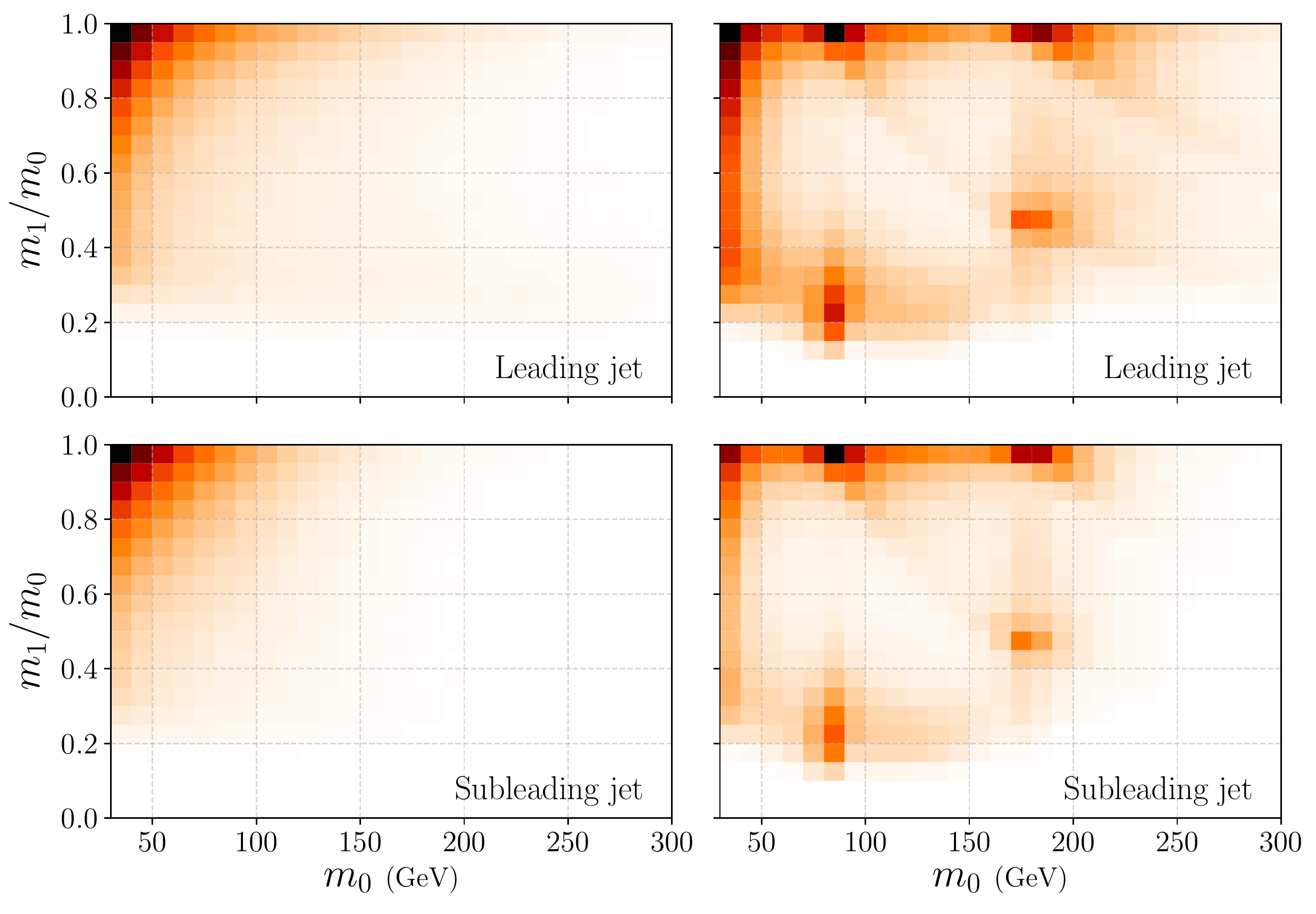}}
  \caption{Distributions of QCD (left) and $t\bar t$ (right) di-jet events in the ($m_{0}$, $m_{1}/m_{0}$) plane. See text for details.\label{fig:tt_m_topics}}
  \vspace{0.5cm}
  \centerline{\includegraphics[scale=0.45]{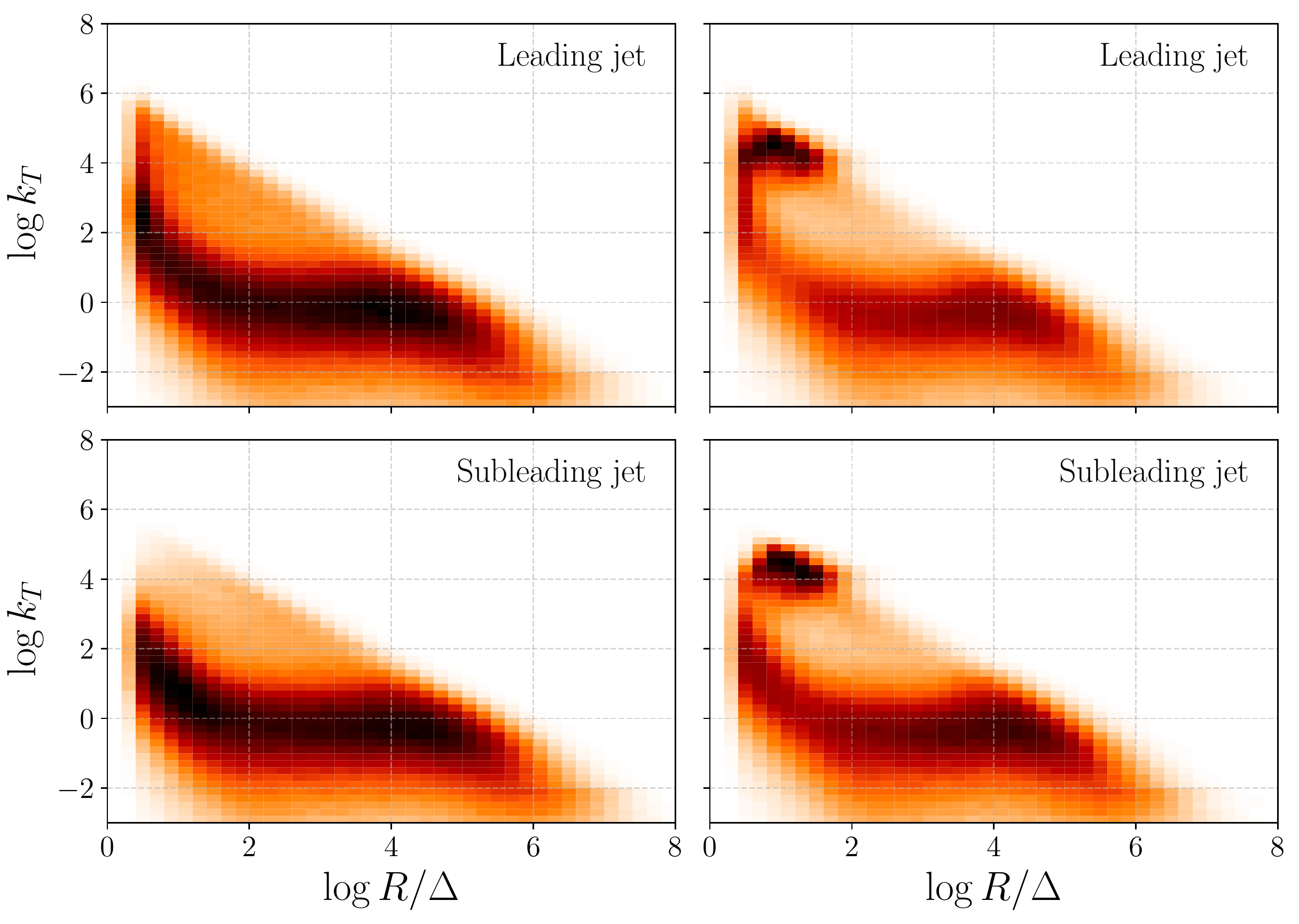}}
  \caption{Distributions of QCD (left) and $t\bar t$ (right) di-jet events in the ($\log k_T$, $\log R/\Delta$) plane. See text for details. \label{fig:tt_l_topics}}
\end{figure}

In Figs.~\ref{fig:tt_m_topics} and~\ref{fig:tt_l_topics} we plot the pure signal ($t\bar{t}$ jets) and background (QCD di-jets) samples in the $(m_{0},m_{1}/m_{0})$ and $(\log R/\Delta,\log k_T)$ planes, respectively.
We see in Fig. \ref{fig:tt_m_topics} that the hard splittings corresponding to the decay of the top quark to the $W$ boson and the decay of the $W$ boson to light jets are clearly discernable.
The top quark decay is indicated by the two clusters (overdensities) of measurements at $m_{0}\simeq 175$ GeV, with the cluster at $m_{1}/m_{0}\simeq1$ being due to the clustering of light radiation around the subjet containing all of the top quark decay products, while the cluster at $m_{1}/m_{0}\simeq m_W/m_t$ corresponds to the splitting that separates the bottom and $W$ subjets from within the top jet.
The decay of the $W$ boson is indicated by the two clusters at $m_{0}\simeq 80$ GeV.
Again the cluster at $m_{1}/m_{0}\simeq 1$ is due to the clustering of soft radiation around the subjet containing the $W$ boson decay products, while the cluster at lower mass drop shows splittings that separate the decay products of the $W$ boson.
The fact that this cluster is at  $m_{1}/m_{0}\simeq 0.2$ does not indicate that the $W$ boson is decaying to a state of mass $\simeq0.2 m_W$, instead this is an artifact of the definition of mass drop $m_1/m_0$ with $m_1>m_2$ ordering.
In mass drop we take $m_{1}$ to be the heaviest of the subjets in the splitting, therefore the distribution of the mass drop is skewed away from zero.
If we instead had plotted $m_{2}/m_{0}$ we would see that this cluster is pushed towards $m_2/m_0 \simeq 0$.
The $(m_{0},m_{1}/m_{0})$ distribution for the background QCD jets is smooth and monotonically decaying at large $m_{0}$ and small $m_{1}/m_{0}$.
In Fig. \ref{fig:tt_l_topics} we see that the splittings corresponding to the hard decays of the top quark and the $W$ boson are indicated by the two overlapping clusters at $\log k_T\simeq 5$ and $\log R/\Delta\simeq 1$.
Apart from the obvious difference in choice of observables here, we should also keep in mind that the actual splittings which pass the cuts here are different than those that pass the cuts in Fig. \ref{fig:tt_m_topics} (see Sec. \ref{sec:jetreps}).
This choice leads to a larger overlap between the background and signal distributions, as seen by the stream of splittings at low $\log k_T$, however there is still a good separation between the features that distinguish the $t\bar{t}$ jets from the QCD background jets.

\subsubsection{A $3$ TeV $W'$ model with a $400$ GeV scalar}\label{sec:bm2}
%%%%%%%%%%%%%%%%%%%%%%%%%%%%%%%%%%%%%%%%%%%%%%%%%%

\noindent The second benchmark is an example of a new physics signature which could be searched for at high-energy colliders using these techniques.
The new physics process is the production of a $3$ TeV $W'$ boson at a collision energy of $13$ TeV, which decays to a SM $W$ boson and a $400$ GeV new physics scalar boson $\phi$.
The scalar boson $\phi$ then subsequently decays to two SM $W$ bosons. The model has been introduced and previously studied in~\cite{Collins:2018epr, Collins:2019jip, Agashe:2018leo}.
For the study in this paper we consider only the hadronic final states of the $W$ bosons.
The mass difference between the $W'$ and its decay products mean that the constituents from the scalar boson and the $W$ will be clustered into a pair of boosted jets, making the jet substructure an important tool for any analysis of these events.
This was first studied in \cite{Collins:2018epr} and used as a benchmark for the unsupervised CWoLa search technique in \cite{Collins:2019jip}.
In the precursor to this paper~\cite{Dillon:2019cqt} this example was also used. 
For this benchmark, in addition to the $p_T$ cut at 400~GeV, events were selected in the di-jet invariant mass window $[2700,3300]$~GeV to encapsulate the peak in the production cross-section of the $W'$ boson.

\begin{figure}[t]
  \centerline{\includegraphics[scale=0.45]{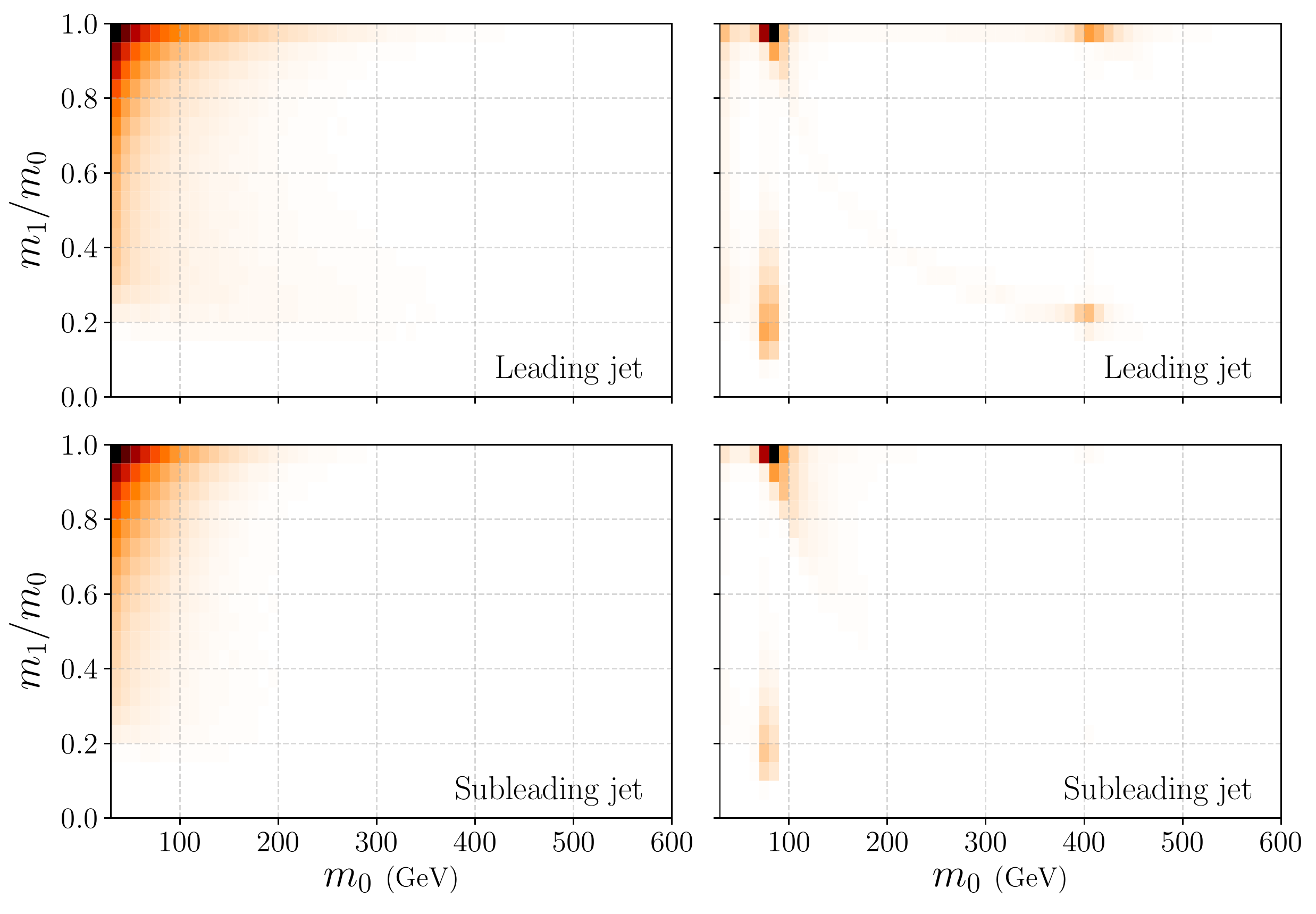}}
  \caption{Distributions of QCD (left) and $W'$ (right) di-jet events in the ($m_{0}$, $m_{1}/m_{0}$) plane. See text for details.  \label{fig:wp_m_topics}}
  \vspace{0.5cm}
  \centerline{\includegraphics[scale=0.45]{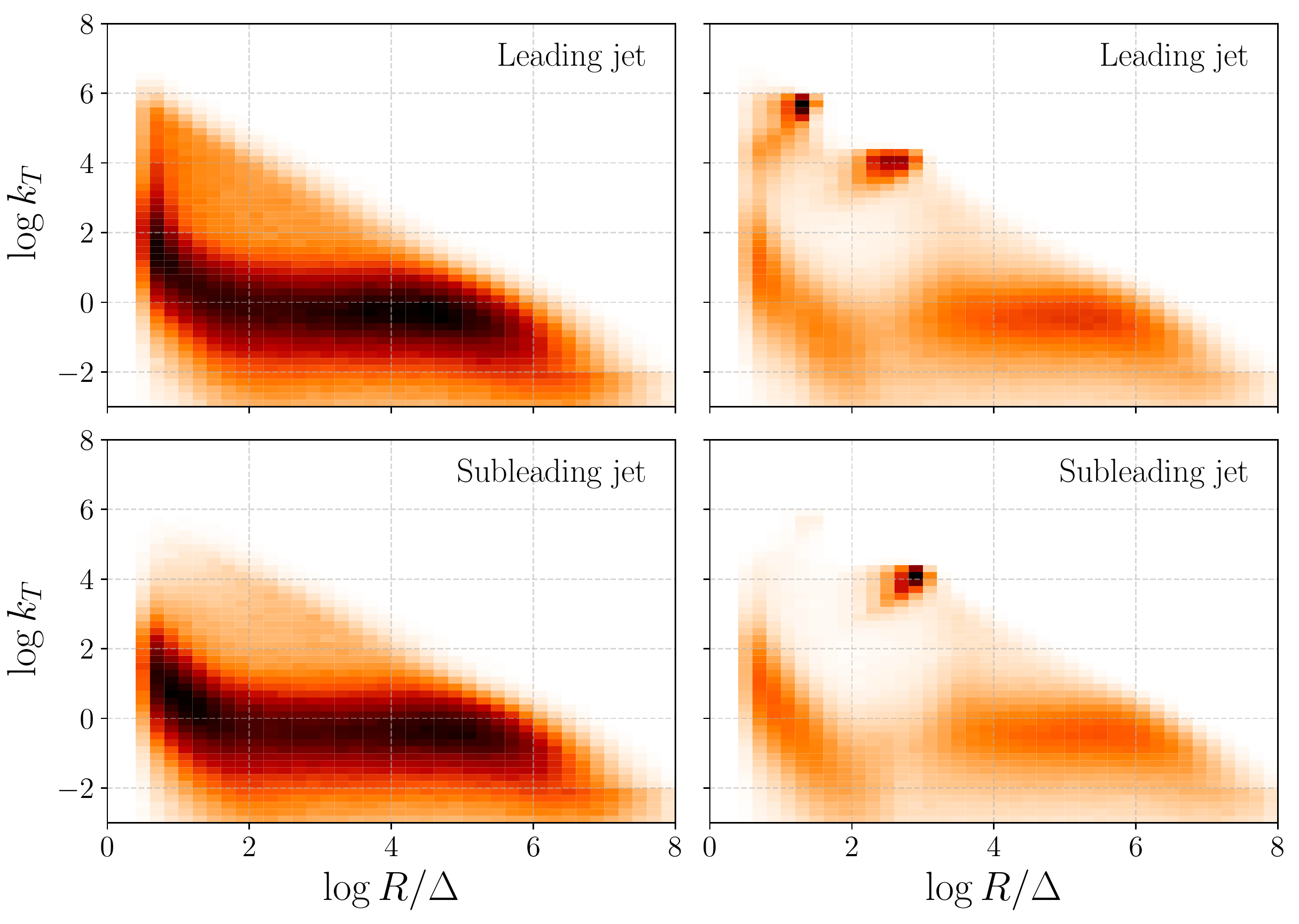}}
  \caption{Distributions of QCD (left) and $W'$ (right) di-jet events in the ($\log k_T$, $\log R/\Delta$) plane. See text for details.   \label{fig:wp_l_topics}}
\end{figure}

In Figs.~\ref{fig:wp_m_topics} and~\ref{fig:wp_l_topics} we plot the pure signal (hadronic $W'$ final states) and background (QCD di-jets) samples in the $(m_{0},m_{1}/m_{0})$ and $(\log R/\Delta,\log k_T)$ planes, respectively.
The origin of the features in these plots is completely analogous to those for $t\bar{t}$ in Figs.~\ref{fig:tt_m_topics} and~\ref{fig:tt_l_topics}.
One important difference to note is that the plots for the leading and subleading signal jets are different, while for $t\bar{t}$ they were equivalent.
This is obviously because here our signal consists of two jets of different origin.
This highlights the importance of including labels for the jets in our representation of the measurements in LDA, in order to properly characterise the signal from the posterior theme distributions.
In Fig.~\ref{fig:wp_m_topics} we can clearly see the clusters corresponding to the decays of the scalar boson in the leading jet at $m_{0}\simeq 400$ GeV.
Of the two clusters at $m_{0}\simeq 400$ GeV the mass drop $m_{1}/m_{0}\simeq 1$ again corresponds to the clustering of soft radiation around the subjet containing all of the scalar boson decay products, while the cluster at $m_{1}/m_{0}\simeq m_{\phi}/m_{W'}$ corresponds to splittings that separate the SM $W$ boson subjets from within the $\phi$ jet.
The clusters corresponding to the decay of the SM $W$ bosons at $m_{0}\simeq 80$ GeV have the exact same features as those in the $t\bar{t}$ case.
The distributions for the background jets in Fig. \ref{fig:wp_l_topics} are similar to the distributions for the background jets in Fig. \ref{fig:tt_l_topics}, as expected.
Interestingly, the distributions for the $W'$ and $t\bar{t}$ signal jets in Fig.'s \ref{fig:wp_l_topics} and \ref{fig:tt_l_topics} are more similar than they are in Fig.'s \ref{fig:wp_m_topics} and \ref{fig:tt_m_topics}.
This is because the observables in the former case measure the $k_T$ and angular separation, rather than the masses of the (sub)jets in the splittings.
Also, the observables are now both binned and displayed on a logarithmic scale, making any differences at large $k_T$ less pronounced.
One obvious difference between the $W'$ and $t\bar{t}$ distributions in Fig.'s \ref{fig:wp_l_topics} and \ref{fig:tt_l_topics} is that the clusters associated with the different hard decays are more distinguishable from each other in the $W'$ case than in the $t\bar{t}$ case.
This is primarily because the mass difference between the scalar boson $\phi$ and the SM $W$ bosons is much larger than the mass difference between the top quark and the SM $W$ bosons.
Another difference is in the amount of soft radiation in the $t\bar{t}$ jets and the $W'$ jets, this is due to the top quark carrying color charge and the $\phi$ boson being color-neutral.
The similarities in the two distributions do however suggest that any classifier selecting events with splittings in the large $k_T$ region may work reasonably well as a generic anti-QCD tagger.

%%%%%%%%%%%%%%%%%%%%%%%%%%%%%%%%%%%%%%%%%%%%%%%%%%
\subsection{Comparing classification power of different observables}\label{sec:classpower}
%%%%%%%%%%%%%%%%%%%%%%%%%%%%%%%%%%%%%%%%%%%%%%%%%%

\begin{figure}[t]
  \centerline{\includegraphics[scale=0.45]{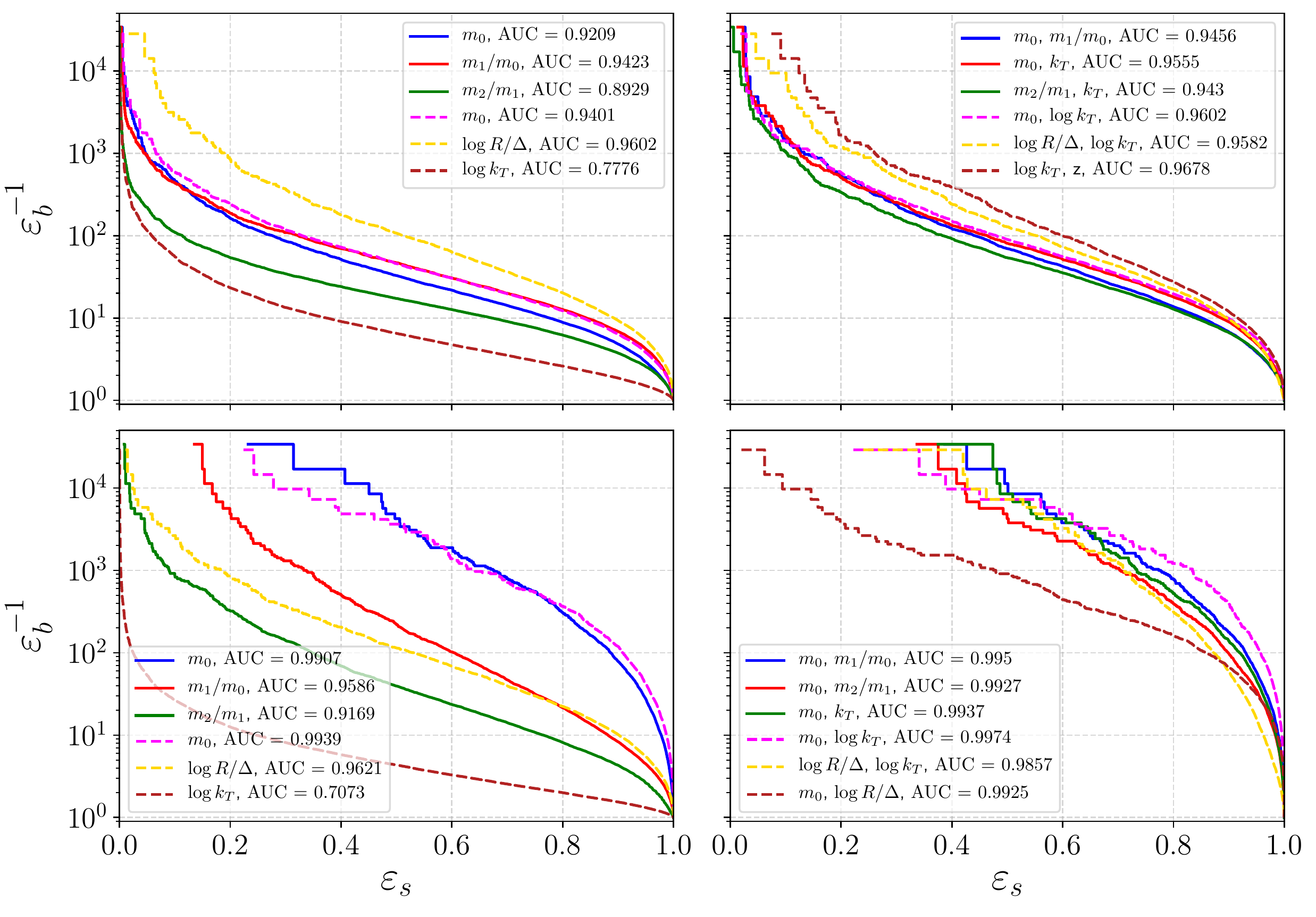}}
  \caption{Classification power of individual observables (left column) and pairs of observables (right column) for both $t\bar{t}$ (top row) and $W'$ (bottom row) signals versus QCD.
We consider both mass basis observables (solid lines), and observables in the primary Lund plane (dashed lines). 
\label{fig:classpower}}
\end{figure}

\noindent There are many possible choices of observables that we could include in our analysis of di-jet events using  LDA.
All of the observables discussed in Sec. \ref{sec:probmodelling} carry some ability to distinguish between signal events and QCD background events, and some observables will be more useful than others depending on what the signal process is.
In this section we study the classification power of each of these observables, and some combinations of them, using a simple binned likelihood classifier.
To construct the binned likelihood classifier we split our signal and background datasets each into `training' and `testing' sets.
We then compile counts of how often each measurement bin occurrs in each of the signal and background training sets, and normalise these to give us a discrete probability distribution for the signal and background samples.
For each event in the testing sets we then compute the likelihood ratio as defined in Eq. \eqref{eq:lr}, except with the $\beta$'s replaced with the binned likelihood multinomials.
The results are summarised in Fig. \ref{fig:classpower}.
First thing to notice is that the observables are in general better at classifying  $W'$ events than  $t\bar{t}$ events, the obvious reason being that the $W'$ signal contains splittings that are very rare in QCD background events, in particular rarer than the splittings in $t\bar{t}$ events.
In the first row we show the classification performance of the observables for the $t\bar{t}$ sample.
The best performing individual observables are $\log R/\Delta$ and $m_{0}$ from the Lund basis.
Note that $m_{0}$ appears twice, once in the mass basis and once in the Lund basis.
The difference in classification power here comes only from the cuts performed on the dataset, because in the mass and Lund bases these cuts differ, as explained in Sec.~\ref{sec:jetreps}.
In combining observables we see that the best performing pair of observables in the Lund basis are $\log k_T$ and $z$.
In the mass basis the best performing pair are $m_{0}$ and $k_T$, however the differences between this pair and others are miniscule.
In the second row we in turn show the analogous plots demonstrating the classification power of the observables for the $W'$ sample.
The results here are different than for $t\bar{t}$, which is not surprising since not only are the masses of the particles produced in the collision different, but also the top quarks are coloured, have spin $1/2$, and therefore produce a very different radiation pattern than the $W'$ decay products (colourless $W$ with spin 1  and $\phi$ with spin 0 ).
The best performing individual observable here is the subjet mass $m_{0}$ in both bases, mass and Lund.
In combining observables we find that the best performing pair of observables in the Lund basis are $m_{0}$ and $\log k_T$, while in the mass basis the best performing pair are $m_{0}$ and $m_{1}/m_{0}$.
Again the differences between these pairs and some of the others are very small.
We do not study combinations of more than two observables, because we find that adding more observables to the best performing pairs does not provide any appreciable difference in classification power of the binned likelihood.
Interestingly however, we have also found that the observables which provide the best performance with the supervised binned likelihood classifier are not necessarily the best to use in an unsupervised analysis based on LDA and VI, which crucially depend on patterns of concurrence of two or more measurements within the same event. 
Therefore in the unsupervised analyses we focus solely on two pairs of observables (i) ($m_{0}$, $m_{1}/m_{0}$), (ii) ($\log k_T$, $\log R/\Delta$), based on their robust performance, good interpretability and since they are already commonly used in jet classification tasks. 
Note that while the classification power of any combination of observables in the supervised binned likelihood does not necessarily indicate the best choice to use in any given analysis, according to the Neyman-Pierson lemma~\cite{Neyman:1933wgr} it does represent an upper bound on the classification power of any unsupervised classifier based on the corresponding themes of these same binned observables, as extracted from LDA. 

%%%%%%%%%%%%%%%%%%%%%%%%%%%%%%%%%%%%%%%%%%%%%%%%%%
\subsection{Measurement co-occurrences}\label{sec:fc}
%%%%%%%%%%%%%%%%%%%%%%%%%%%%%%%%%%%%%%%%%%%%%%%%%%

\noindent When introducing LDA and VI in Sec.~\ref{sec:probmodelling} we first encountered the importance of measurement co-occurrence in individual events.
Certain measurements within individual events must exhibit a pattern of co-occurrences in order for the inference algorithm to recognise and extract the corresponding theme distributions. In other words, VI is unable to extract any information from unique measurements $o_{j,i}$ appearing only once in the dataset. In this section we explore how the measurement co-occurrences  in a dataset vary with the choice of observables and their binning, highlighting the importance of the data representation in the construction of the unsupervised classification strategy. We demonstrate this on the example of the $W'$ model, while we have checked that the $t\bar{t}$ example exhibits analogous behaviour.

\begin{figure}[t]
  \centerline{\includegraphics[scale=0.45]{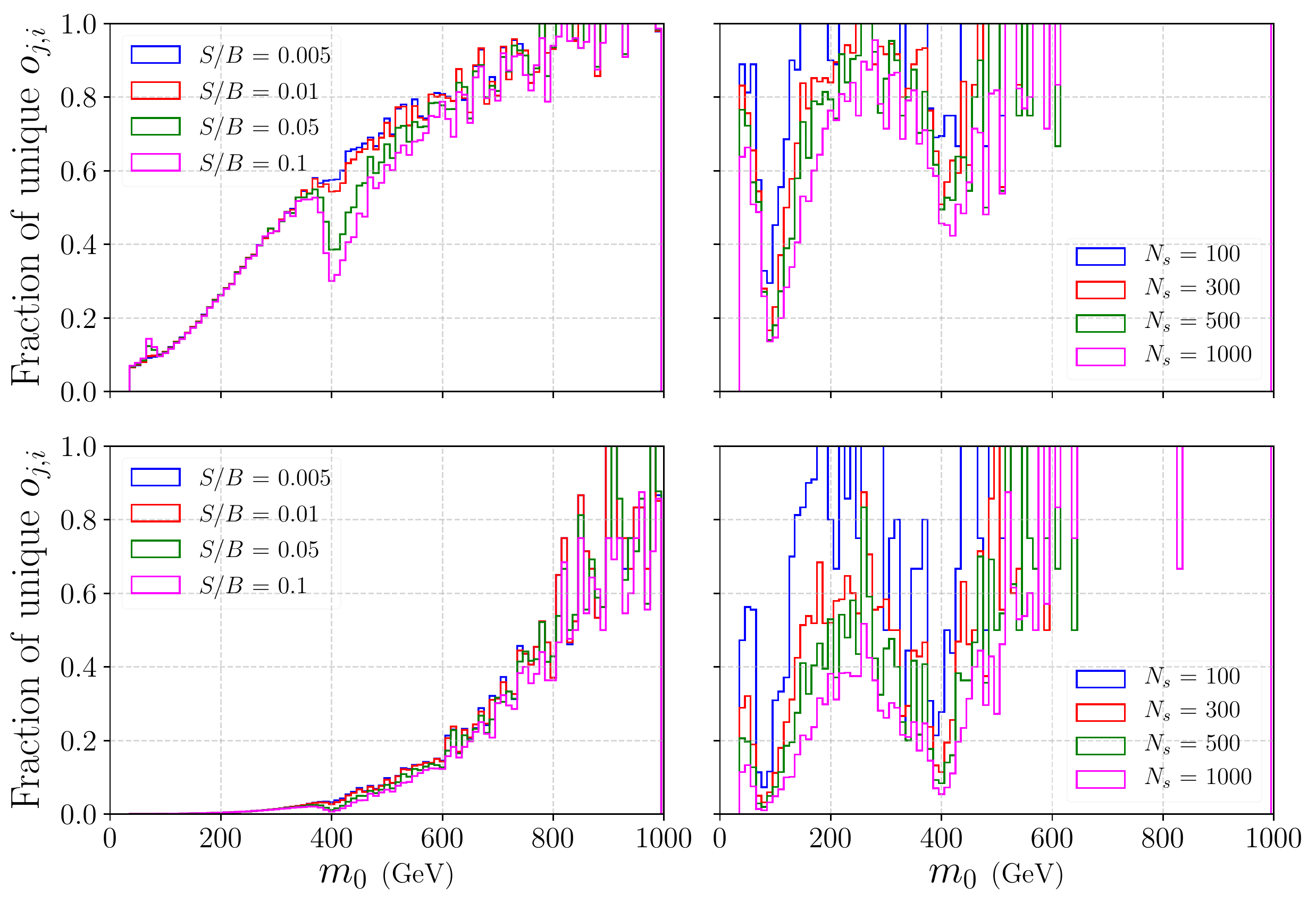}}
  \caption{Fraction of unique measurements in $W'$ event samples, using all the mass basis observables ($m_{0}$, $m_{1}/m_{0}$, $m_{2}/m_{1}$, $k_T$, $\cos\theta$) (top line) and only the pair ($m_{0}$, $m_{1}/m_{0}$) (bottom line). On the left we show samples of $9\!\times\! 10^4$ events consisting of different fractions of mixed signal ($W'$) and background (QCD) events, while on the right we show the results for different numbers of pure signal  ($W'$)  events.  \label{fig:cooccur1}}
 \vspace{0.5cm}
  \centerline{\includegraphics[scale=0.45]{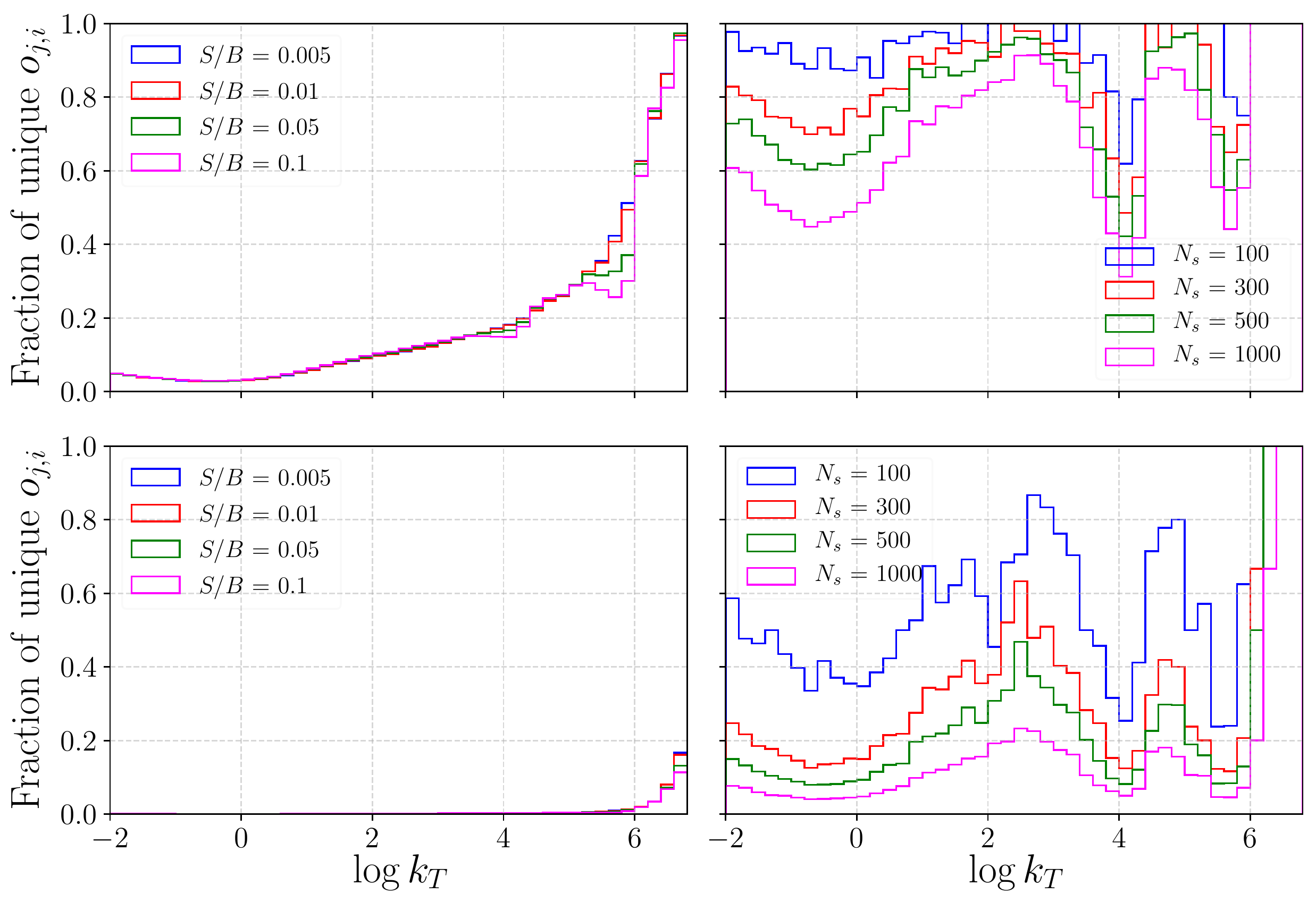}}
  \caption{Fraction of unique measurements in $W'$ event samples, using all the Lund basis observables ($m_{0}$, $\log R/\Delta$, $\log k_T$, $z$, $\log R/\kappa$, $\psi$) (top line) and only the pair  ($\log k_T$, $\log R/\Delta$) (bottom line). On the left we show samples of $9\!\times\! 10^4$ events consisting of different fractions of mixed signal ($W'$) and background (QCD) events, while on the right we show the results for different numbers of pure signal  ($W'$)  events. \label{fig:cooccur2} }
\end{figure}

We quantify the measurement co-occurrences by calculating the number of unique $o_{j,i}$ per bin of one of the observables, marginalising over the rest, and dividing this by the total number of measurements in that bin.
The lower this `fraction of unique $o_{j,i}$' is the stronger the co-occurrences are, and the easier it will be for the VI algorithm to extract themes accurately describing the underlying structure of the events. In the upper row of Fig. \ref{fig:cooccur1} we show how the co-occurrences in the mixed $W'$-QCD sample with observables ($m_{0}$, ${m_{1}}/{m_{0}}$, ${m_{2}}/{m_{1}}$, $k_T$, $\cos\theta$) vary per $m_{0}$ bin.
On the left hand side we do this for a mixed sample of $9\!\times\! 10^4$ signal and background events with varying S/B, while on the right hand side we do it for varying amounts of pure signal events ($100$, $500$, and $1000$).
We focus on such small numbers of signal events because we are interested in finding rare signals, in which the co-occurrences will inevitably be less apparent. Conversely, in a sample containing a large fraction of signal events the structure of the signal events would be more easily uncovered due to the strong co-occurrences between the measurements.
As expected, the co-occurrences are strongest at $m_0\simeq m_W$ and $m_0 \simeq m_\phi$, since the signal events are more likely to contain splittings with these masses (see Fig. \ref{fig:wp_m_topics}).
However, as discussed in the previous subsection, including more than two of these observables in the analysis does not significantly increase the classification power of the binned likelihood classifier. At the same time we demonstrate in the second row how restricting the analysis to including just one such pair ($m_{0}$, $m_{1}/m_{0}$) can drastically increase the strength of the co-occurrences in the event sample.
This provides further justification for including no more than two observables in the LDA analysis.
In Fig. \ref{fig:cooccur2} we display the same information for the Lund observables where we measure the co-occurrences as a function of $\log k_T$.
We see again that the co-occurrences are strongest at the points where the signal features are most pronounced (see Fig. \ref{fig:wp_l_topics}), and that by restricting the observables used at each splitting we can increase the frequency of these co-occurrences significantly.

\begin{figure}[t]
  \centerline{\includegraphics[scale=0.45]{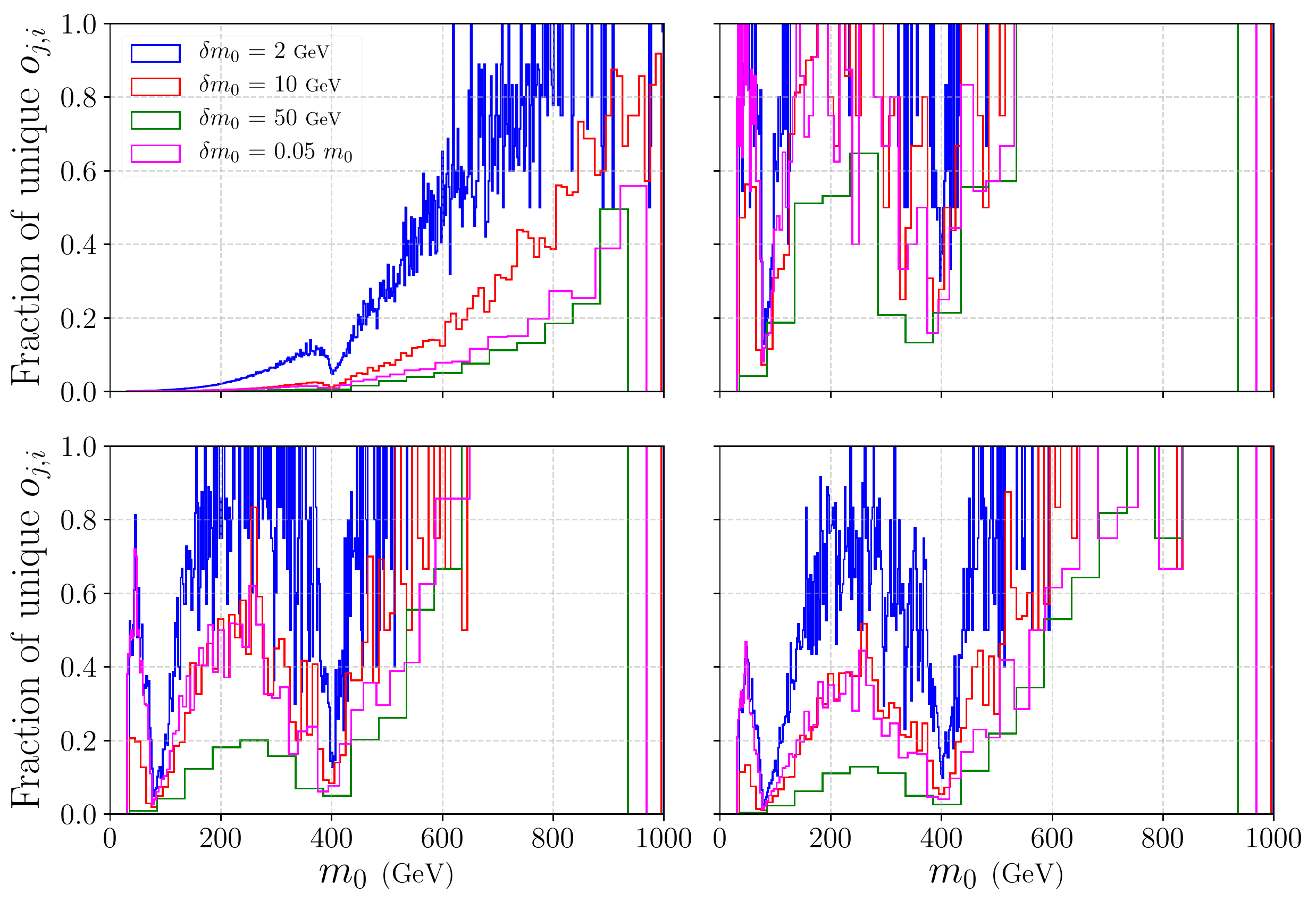}
  }
  \caption{Fraction of unique measurements in $W'$ event samples, using the mass basis observables ($m_{0}$, $m_{1}/m_{0}$) for different choices of $m_0$ binning.  From top-left to bottom-right:  results for a mixed sample of $9\!\times\! 10^4$ events with ${\rm S/B}=5\%$ and four different choices of bin sizes, followed by results for 100, 500 and 1000 pure signal events at various bins sizes of $\delta m_0 = 2$~GeV, $10$~GeV $50$~GeV, and $0.05\times m_{0}$.  \label{fig:cooccur3}}
\end{figure}

Before moving on we examine another handle we have on increasing co-occurrences in the event sample, that is by varying the binning used for each of the observables.
To do this we keep with the $W'$ sample and focus on just the pair of ($m_{0}$, $m_{1}/m_{0}$) observables. The results are summarised in Fig. \ref{fig:cooccur3}.
We note that some of the bin sizes used in this plot would be impossible to use in practice due to the finite experimental resolution, however they still serve as useful examples to demonstrate the potential effects of varying bin sizes in the analysis. 
In the upper left plot we show the co-occurrences for the whole mixed sample with S/B=$5\%$ and four different choices of bin sizes.
In each of the other three plots we then show the co-occurrences for different numbers of signal events (again $100$, $500$, and $1000$) and varying bin sizes.
As expected, larger bin sizes result in stronger co-occurrences, however the size of this effect is not as large as the effect of removing observables from the analysis completely.
For example, in all cases the strength of the co-occurrences at $m_0\simeq m_W$ is almost the same for all choices of the binning.
The effect due to different bin sizes is more clearly seen away from these areas of strongest co-occurrence, where larger bin sizes result in stronger co-occurrences across the whole $m_{0}$ range. In particular, this may aid in better modelling of the signal and background distributions away from $m_{0}\simeq m_W$ and $m_0 \simeq m_\phi$.
On the other hand, increasing the bin size will also make the signal features less pronounced, potentially reducing the classification power in the same way as a binned likelihood classifier becomes worse and worse approximation to the Neyman-Pearson un-binned likelihood. Therefore there is trade-off here between potential classification power and the ability of VI to extract optimal theme distributions from the data. In particular we find that the constant $\delta m_0 = 10$ GeV bin size provides the best trade-off for the examples satudied here and is also in practice close to the variable binning $\delta m_0 = 0.05 m_0$ mimicking the typical (energy) resolution of modern particle detector calorimeters.

%%%%%%%%%%%%%%%%%%%%%%%%%%%%%%%%%%%%%%%%%%%%%%%%%%
%
\section{Unsupervised learning with LDA}\label{sec:ldaresults}
%
%%%%%%%%%%%%%%%%%%%%%%%%%%%%%%%%%%%%%%%%%%%%%%%%%%

\noindent As our main result we present two applications of the technique outlined in the preceding sections.
Using the two benchmark examples discussed in Sec. \ref{sec:bm1} and Sec. \ref{sec:bm2} we construct various mixed event samples, i.e. mixtures of background and signal events.
For the boosted $t\bar{t}$ example we construct mixed samples with $9\times 10^4$ events, and with S/B ratios: $1$, $0.5$, $0.1$, $0.05$, and $0.01$. 
With the $W'$ example we construct mixed samples with $9\times 10^4$ events, and with S/B ratios: $0.1$, $0.05$, $0.025$, $0.01$, and $0.005$.  
We also include pure background samples (S/B=0) to demonstrate what the output of LDA looks like with no signal events present.
We focus more on lower S/B ratios for the $W'$ events than we do for $t\bar{t}$ events case because ultimately we are interested in uncovering rare new physics signals in the data.
For each benchmark we construct two separate sets of mixed samples, one using the mass basis observables ($m_{0}$, $m_{1}/m_{0}$) and one using the (primary) Lund plane ($\log k_T$, $\log R/\Delta$), as outlined in Sec. \ref{sec:classpower}.
For each mixed sample, $12$ in total, we train LDA models with different Dirichlet parameters, extract the themes, and use them to cluster/classify the events in the sample.
We perform an extensive scan over the Dirichlet parameters,  training $961$ models  for each mixed sample, scanning over the ($\rho$,$\Sigma$) parameter space in the ranges $-3\leq \log_{10}\rho\leq 0$ and $0\leq\Sigma\leq3$ with resolution $\delta \log_{10}\rho=0.1$ and $\delta\Sigma=0.1$, respectively.
For each of these mixed samples we plot the inverse perplexity ($\mathcal P^{-1}$) calculated over the whole sample, as well as the AUC, and the inverse mistag at fixed efficiency, both calculated on separate pure signal and background samples.
In particular, we are interested in how the inverse perplexity, which can be computed from unlabelled data alone, is correlated with the performance of the classifier (AUC and $\epsilon_b^{-1}\left(\epsilon_s=0.5\right)$), which is inaccessible in absence of labelled data. We also consider how both the inverse perplexity and performance behave in different regions of the Dirichlet parameter space, as discussed in Sec. \ref{sec:lda}.

%%%%%%%%%%%%%%%%%%%%%%%%%%%%%%%%%%%%%%%%%%%%%%%%%%
\subsection{Unsupervised classification of boosted $t\bar{t}$ production}
\label{sec:ttscans}
%%%%%%%%%%%%%%%%%%%%%%%%%%%%%%%%%%%%%%%%%%%%%%%%%%

\begin{figure}[t]
  \centerline{\includegraphics[scale=0.4]{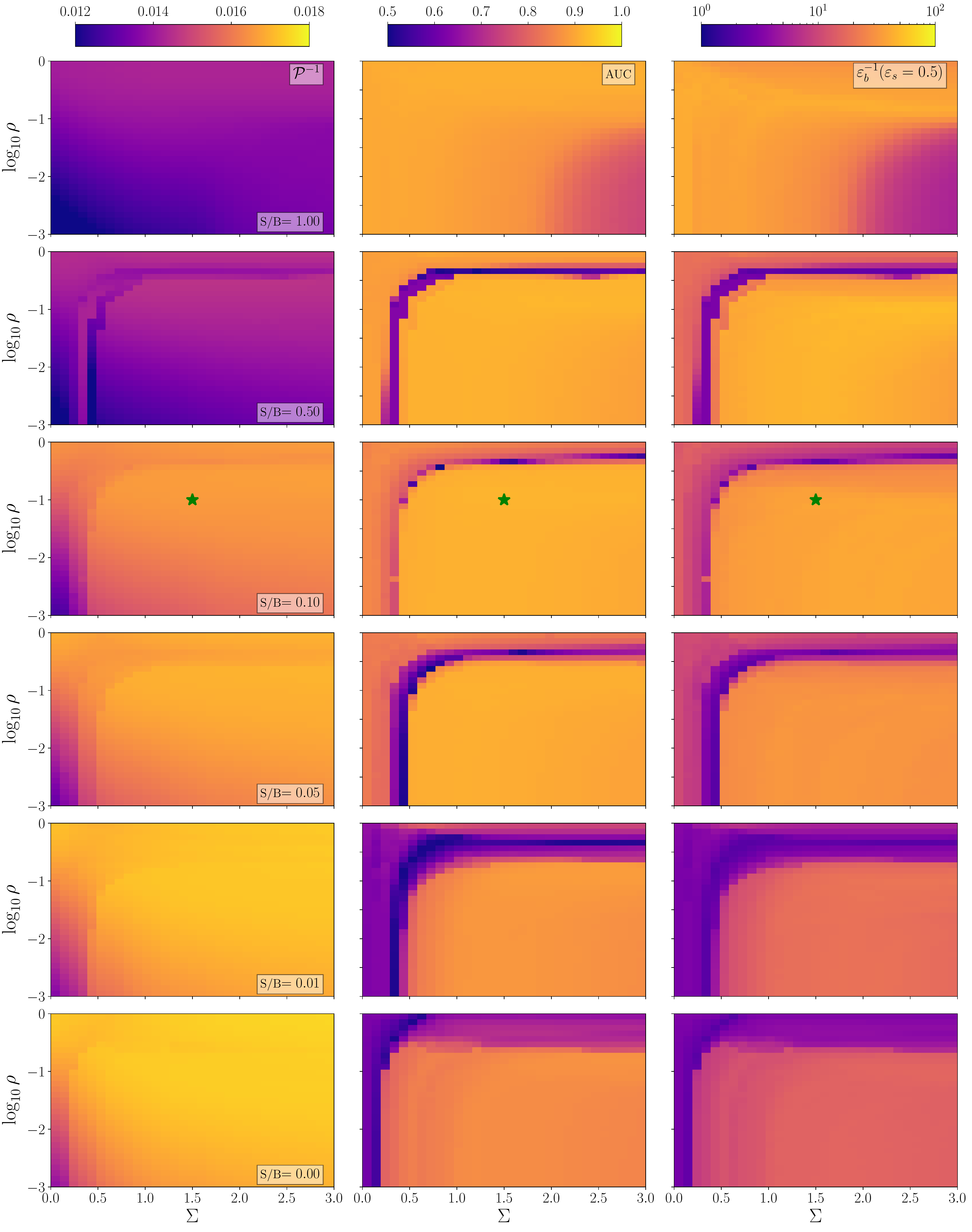}
  }
  \caption{Results of LDA models in the $(\rho,\Sigma)$ parameter-space trained on samples of mixed $t\bar{t}$ and QCD events using mass basis observables $m_{0}$ and $m_{1}/m_{0}$,  with different S/B ratios (one per row).  Each row contains plots of perplexity, AUC, and inverse mis-tag rate at fixed efficiency (see text for details).
The green star indicates the model used to plot the theme distributions in Fig.~\ref{fig:tt_m_theme_extracted}.   \label{fig:tt_m_scan}}
\end{figure}

\begin{figure}[t]
  \centerline{\includegraphics[scale=0.4]{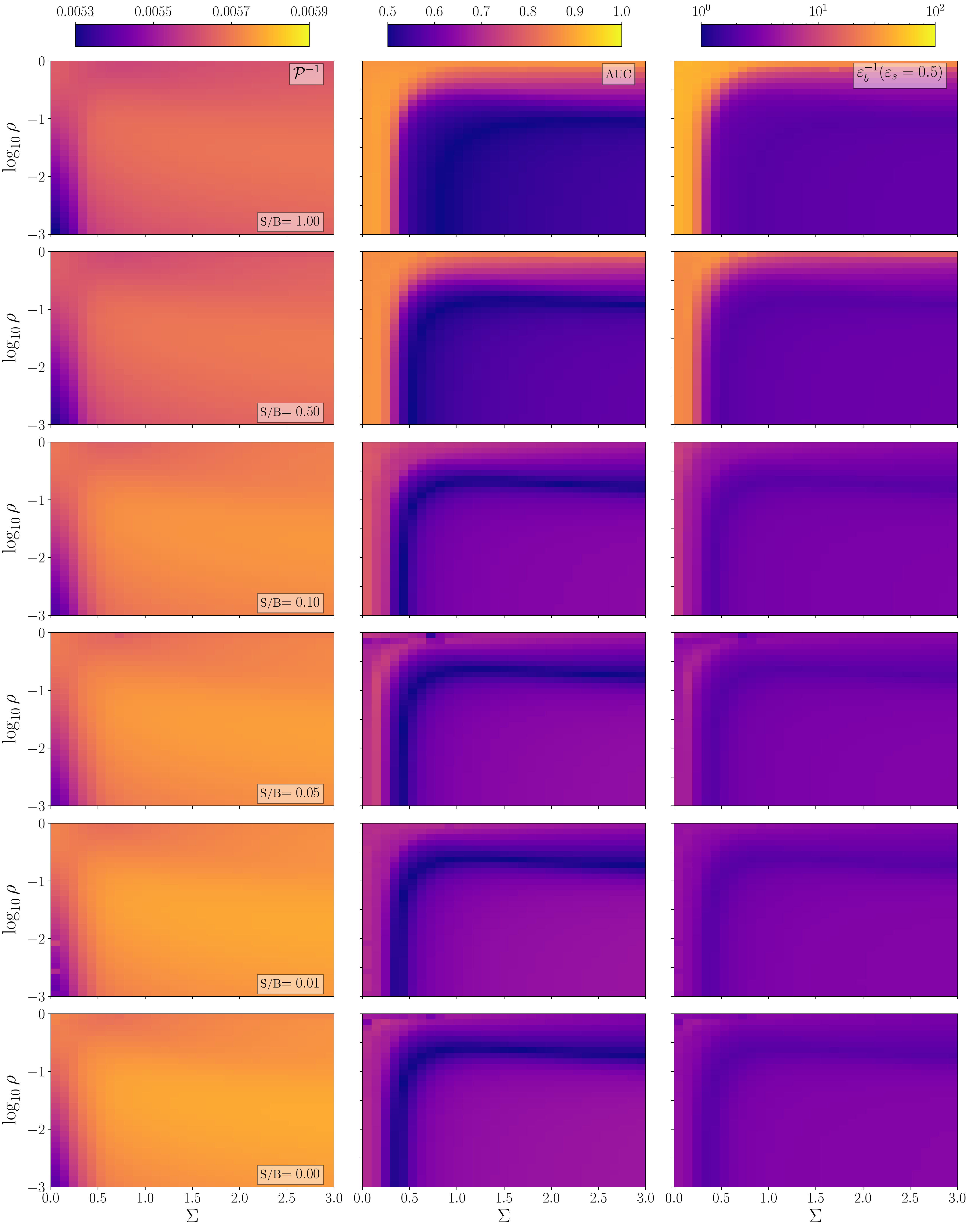}
  }
  \caption{Results of LDA models in the $(\rho,\Sigma)$ parameter-space trained on samples of mixed $t\bar{t}$ and QCD events using Lund basis observables $\log k_T$ and $\log R/\Delta$,  with different S/B ratios (one per row).  Each row contains plots of perplexity, AUC, and inverse mis-tag rate at fixed efficiency.  See text for details.  \label{fig:tt_l_scan}}
\end{figure}

\noindent Starting with the boosted $t\bar t$ scans using the ($m_{0}$, $m_{1}/m_{0}$) observables in Fig. \ref{fig:tt_m_scan}, the first obvious trend we see is that the performance of the LDA classifiers is reduced as the number of signal events in the sample is decreased.
This occurs because with less signal events it becomes more difficult for the algorithm to extract an accurate description of the signal in terms of the themes.
Thus when we construct the likelihood ratio classifier as in Eq. \eqref{eq:lr}, the performance is reduced for smaller S/B.
In a full search strategy this would make it harder to isolate low S/B signals.
This is a universal trend that we also see in Fig.'s \ref{fig:tt_l_scan}, \ref{fig:wp_m_scan}, and \ref{fig:wp_l_scan}.
For ${\rm S/B}=1$ the best performing models tend to be in the regions with larger $\rho$ and smaller $\Sigma$.
This intuitively makes sense, as the larger $\rho$ implies that the algorithm attempts to extract themes under the assumption that the different types of events making up the mixed sample occur in comparable proportions.
Also, the smaller $\Sigma$ ($\leq 1$) means that the prior over the theme proportions is bi-modal (see red contours Fig. \ref{Beta_distribution}), and so events are assumed to be mostly composed of one dominant theme. 
We see some correlation between the perplexity and performance here, although because the performance is good in most of the parameter space for ${\rm S/B} =1$ there is not much to note.
As we lower the S/B we see immediately, even at ${\rm  S/B} =0.5$, that a ridge-like feature forms in the same region in both the perplexity and performance plots.
This persists all the way to ${\rm S/B} =0$ and seems to be related to the transition with the prior moving from a uni-modal to a bi-modal shape.
We see that the performance of the classifiers is better on one side of the ridge than the other, with the performance being better at lower, rather than larger, $\rho$ as the S/B is lowered.
Regions with large inverse perplexity do seem to indicate regions where the classifier is better at identifying the signal, with one exception.
In the $\rho,\Sigma\rightarrow0$ limit the LDA model tends to a mixture model where only a single theme is relevant.
In this region the inverse perplexity invariably reduces, while the performance of the classifier tends to flatten out. This result also shows that mixed membership models preform better at describing the data compated to mixture models.
Finally we note that the presence or absence of a signal cannot be inferred by looking at the inverse perplexity plot alone. For very low S/B (including ${\rm S/B} =0$), LDA is unable to learn signal features effectively and (depending on the priors) learns to predominantly isolate (rare) co-occurance patterns in the background data. The resulting classifiers in this case (at small $\rho \lesssim 0.1$ and moderate $\Sigma \sim \mathcal O(1)$) are effectively anti-QCD taggers.

In Fig. \ref{fig:tt_l_scan} we show in turn the results of the S/B and ($\rho$,$\Sigma$) scan for $t\bar{t}$ using the pair of ($\log k_T$, $\log R/\Delta$) observables.
We see a very similar general behaviour as in the mass basis example: similar correlation between perplexity and performance, with similar behaviour at $\rho,\Sigma\rightarrow0$ and a similar ridge-like feature forming at $\Sigma<1$.
Albeit the performance of the classifiers here is generally worse than in the ($m_{0}$, $m_{1}/m_{0}$) case in Fig. \ref{fig:tt_m_scan}.
We also see that the AUC even slightly improves for lower S/B when $\rho$ is small and $\Sigma$ is large.
This is again the effect of anti-QCD tagging, where the algorithm learns the distribution describing the QCD background quite well but does not learn the signal distribution. Sometimes this is enough to see a classification performance which is slightly better than random.

\begin{figure}[t]
  \centerline{\includegraphics[scale=0.45]{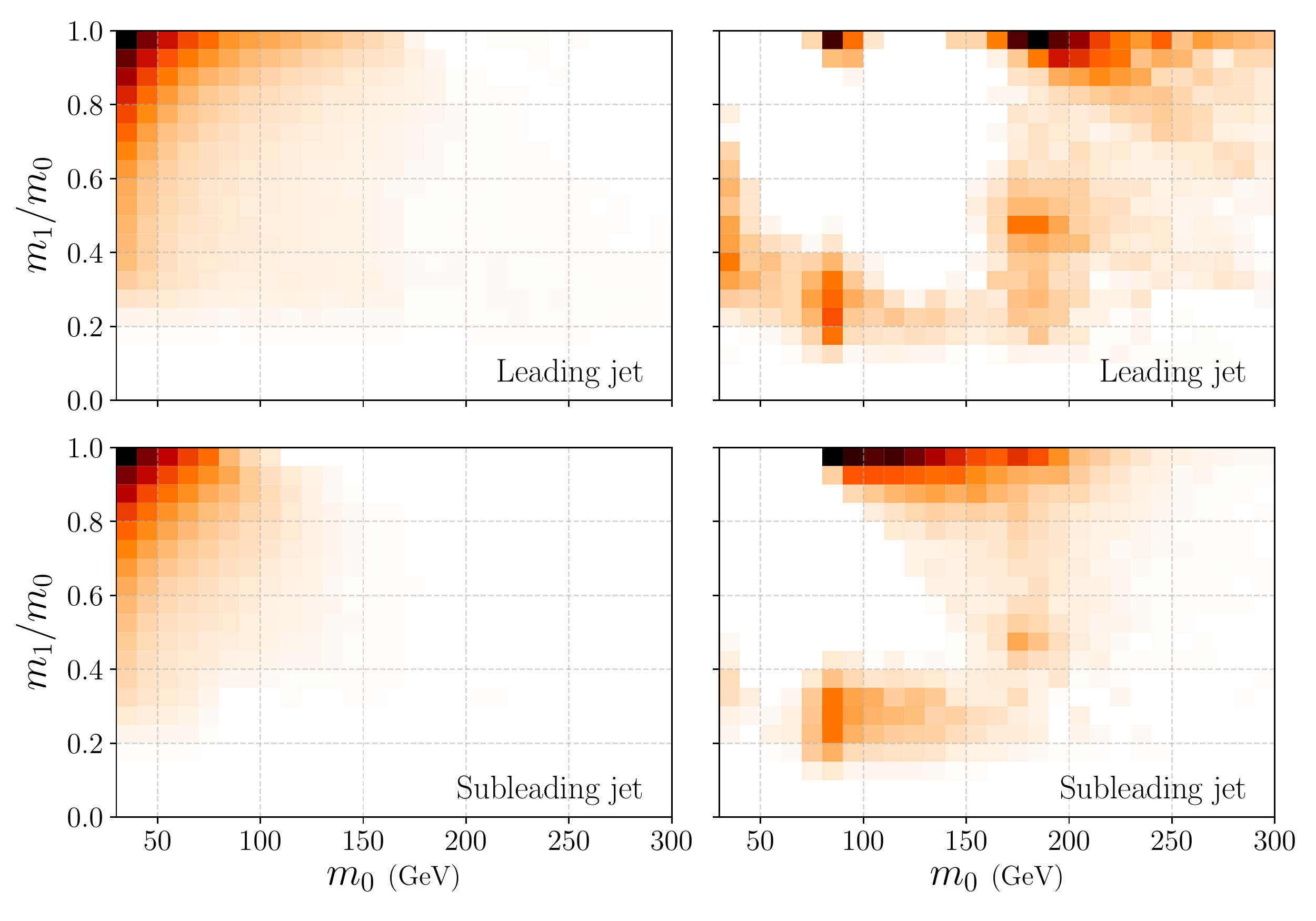}
  }
  \caption{The LDA extracted theme 1 (left) and theme 2 (right) distributions for the leading (upper plots) and subleading (lower plots) jets obtained on a mixed $t\bar{t}$  / QCD sample with S/B at $10\%$, where only the $m_{0}$ and $m_{1}/m_{0}$ observables were used. Shown are results for the model with priors $\rho=0.1$ and $\Sigma=1.5$ which yield the biggest inverse perplexity.
  See text for details.   \label{fig:tt_m_theme_extracted}}
\end{figure}

As an example of the themes that are extracted using this technique we select a model from the ($m_{0}$, $m_{1}/m_{0}$) scan with S/B$=0.1$.
It is important to emphasise that when using these techniques in practice one does not in general have access to pure distributions with which to gauge the classification performance, so the perplexity will be an important statistic to judge which prior parameters best describe the data (and in turn allow for the best classification performance).
Therefore we have selected the model with $\rho=0.1$ and $\Sigma=1.5$, since this model exhibits a large inverse perplexity.
The corresponding learned model themes are shown in Fig. \ref{fig:tt_m_theme_extracted}.
These extracted latent distributions are remarkably similar to the pure distributions of the $t\bar{t}$ event samples shown in Fig. \ref{fig:tt_m_topics}.
Remember that there is a direct expected correlation between LDA themes that are similar to the pure signal and background distributions and the performance of the corresponding theme-based likelihood classifier between signal and background events.
Despite the similarities, there is one important difference between the theme distributions in Fig. \ref{fig:tt_m_theme_extracted} and pure sample distributions in Fig. \ref{fig:tt_m_topics}.
That is the soft QCD splittings (at $m_0\to 0$, $m_1/m_0\to 1$) present in both background and signal events in Fig. \ref{fig:tt_m_topics} are not present in one of the extracted themes. 
This is precisely due to the mixed-membership nature of LDA and is related to the anti-QCD part of the classifier.
The model was able to identify the smooth QCD distribution peaking towards $m_0\to 0$, $m_1/m_0\to 1$ as a theme. While it was able to recognise these same features both in background as well as in signal events, in the later case it  also picked up on co-occouring hard splittings corresponding to the $t$ and $\bar t$ decays and isolated them to a separate theme. 

\clearpage

\clearpage

%%%%%%%%%%%%%%%%%%%%%%%%%%%%%%%%%%%%%%%%%%%%%%%%%%
\subsection{Unsupervised characterisation of new physics}
\label{sec:wpscans}
%%%%%%%%%%%%%%%%%%%%%%%%%%%%%%%%%%%%%%%%%%%%%%%%%%

\begin{figure}[t]
  \centerline{\includegraphics[scale=0.4]{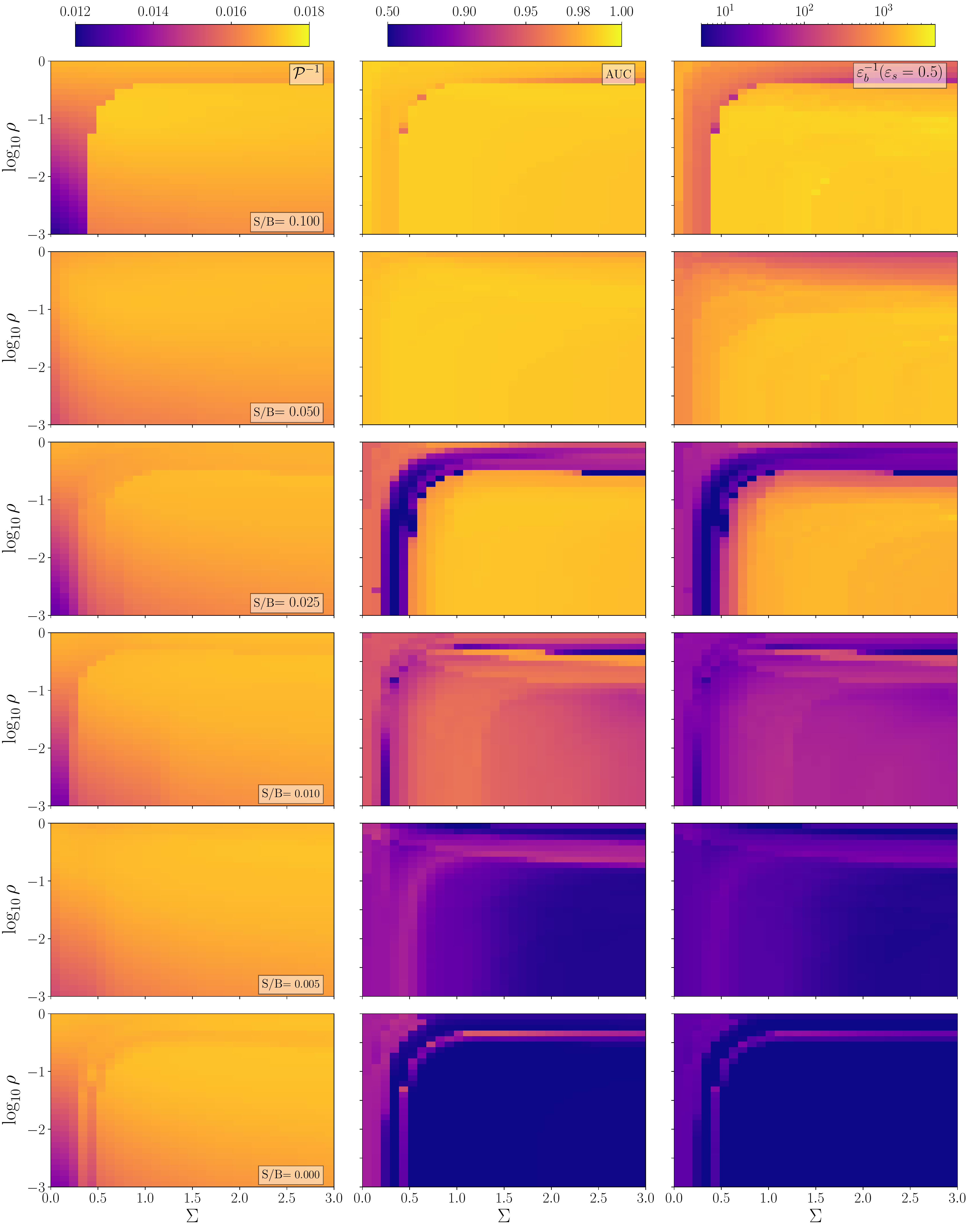}
  }
  \caption{Results of LDA models in the $(\rho,\Sigma)$ parameter-space trained on samples of mixed $W'$ and QCD events using mass basis observables $m_{0}$ and $m_{1}/m_{0}$,  with different S/B ratios (one per row).  Each row contains plots of perplexity, AUC, and inverse mis-tag rate at fixed efficiency.  See text for details.   \label{fig:wp_m_scan}}
\end{figure}

\begin{figure}[t]
  \centerline{\includegraphics[scale=0.4]{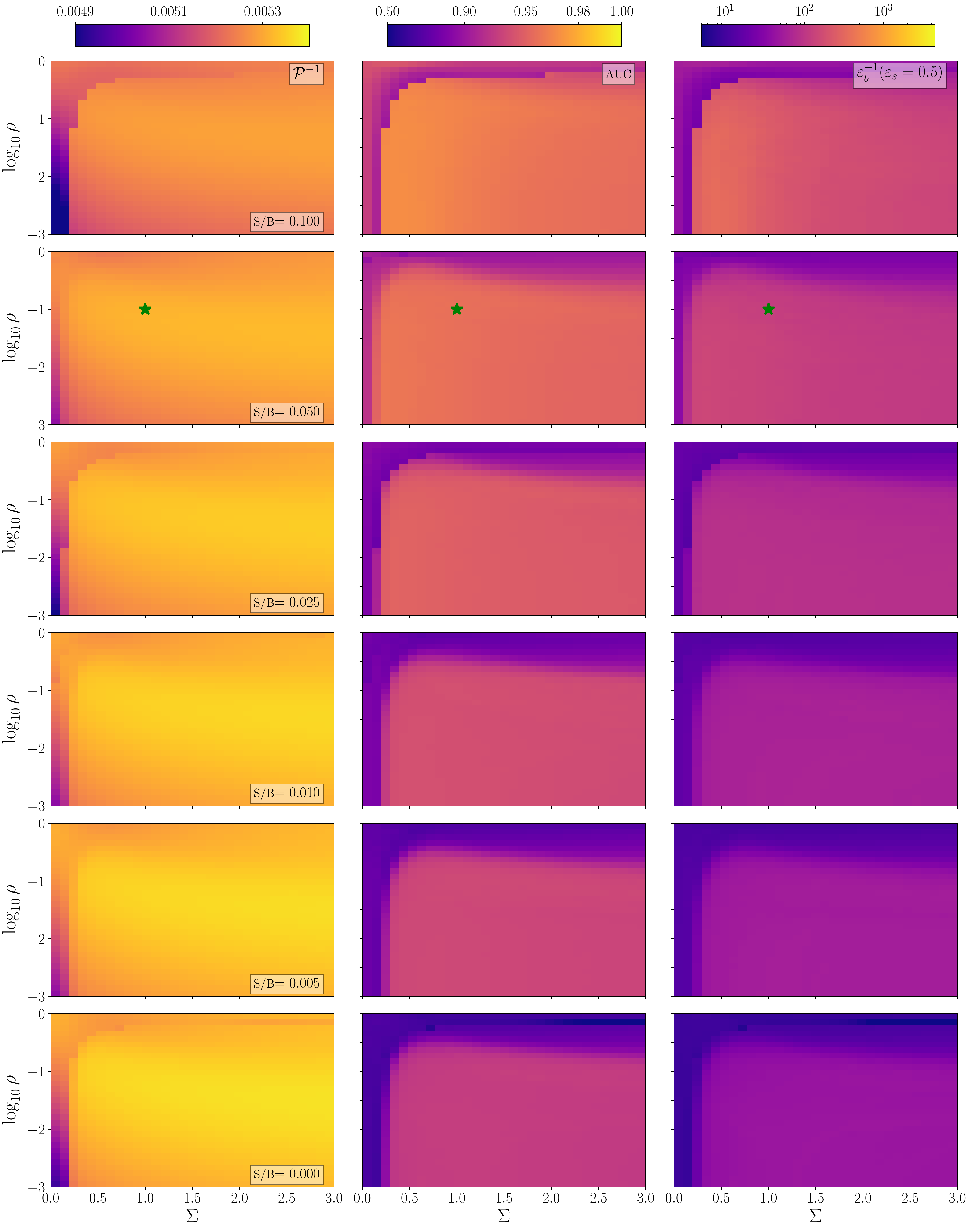}
  }
  \caption{Results of LDA models in the $(\rho,\Sigma)$ parameter-space trained on samples of mixed $W'$ and QCD events using Lund basis observables $\log k_T$ and $\log R/\Delta$,  with different S/B ratios (one per row).  Each row contains plots of perplexity, AUC, and inverse mis-tag rate at fixed efficiency (see text for details).
The green star indicates the model used to plot the theme distributions in Fig.~\ref{fig:wp_l_theme_extracted}.  \label{fig:wp_l_scan}}
\end{figure}

\begin{figure}[t!]
  \centerline{\includegraphics[scale=0.45]{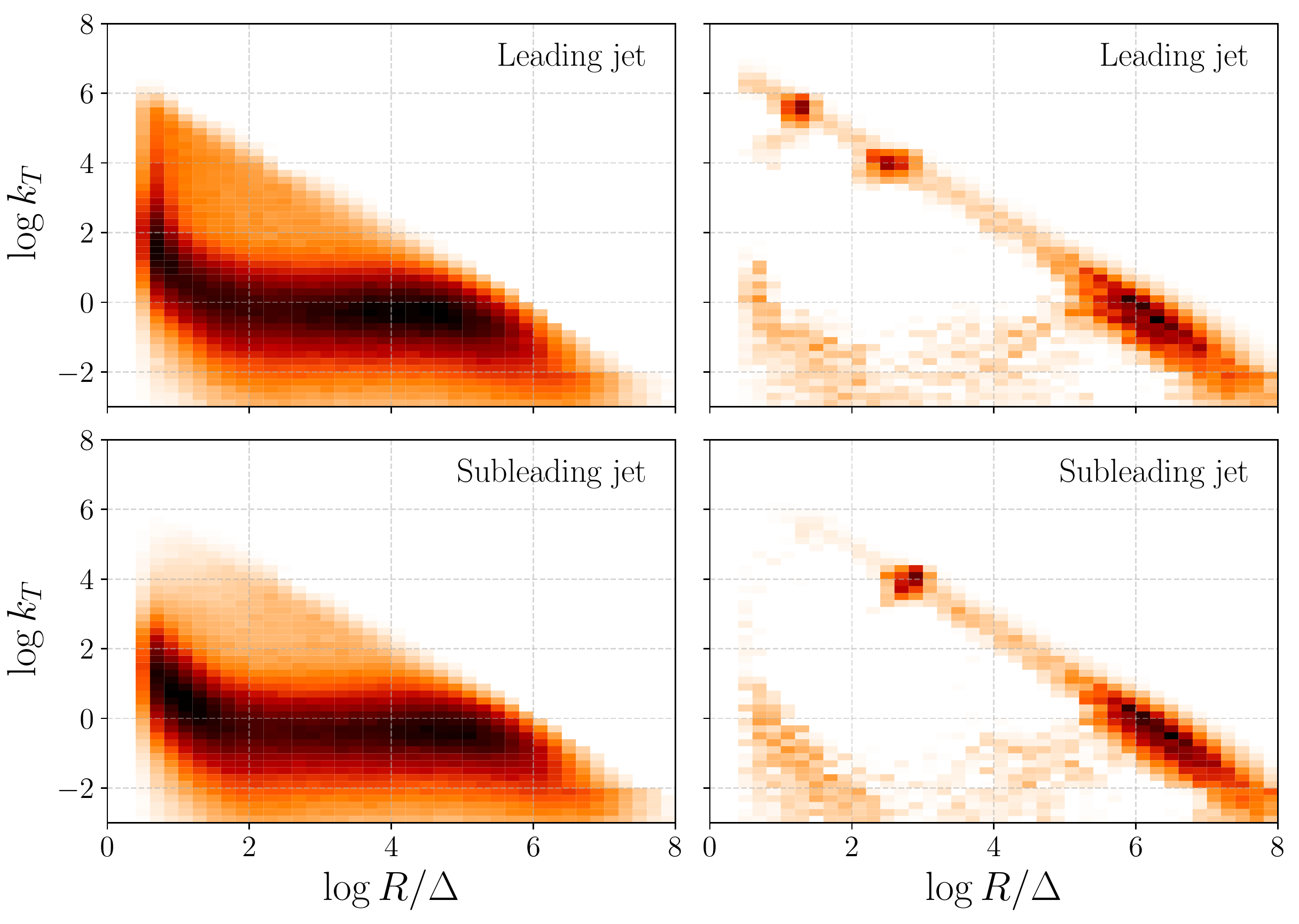}
  }
  \caption{The LDA extracted theme 1 (left) and theme 2 (right) distributions for the leading (upper plots) and subleading (lower plots) jets obtained on a mixed $W'$  / QCD sample with S/B at $5\%$, where only the $\log k_T$ and $\log R/\Delta$ observables were used. Shown are results for the model with priors {$\rho=0.1$ and $\Sigma=1$} which yields the biggest inverse perplexity.
  See text for details.  \label{fig:wp_l_theme_extracted}}
  \end{figure}

\noindent Moving on to the new physics benchmark $pp\rightarrow W'\rightarrow \phi W,~\phi\rightarrow WW$, with $m_W=3$ TeV and $m_{\phi}=400$ GeV,
the scan plots in Figs.~\ref{fig:wp_m_scan} and~\ref{fig:wp_l_scan} are analogous to those for $t\bar{t}$, with different S/B's.
The same features we have seen for $t\bar{t}$ persist here: the ridge-like structure at $\rho\rightarrow0$ and $\Sigma<0.5$, the correlation between perplexity and classifier performance, and the dip in inverse perplexity at $\rho,\Sigma\rightarrow0$.
Clearly the performance of the classifier degrades as the S/B in the sample is reduced, but we see that the performance for the mass basis observables in Fig. \ref{fig:wp_m_scan} remains acceptable down to S/B$=0.01$, while at S/B=$0.005$ it seems that the algorithm finds it difficult to robustly extract a reliable description of the signal.
Note that even in the case with no signal there are still some features in the perplexity plot, and the performance is quite a bit better than a random tagger in some cases.
This is again due to the anti-QCD tagging effect, which is stronger here than in the $t\bar{t}$ benchmark because the defining features of the $W'$ events in this representation have a smaller overlap with the QCD background.
The performance of the Lund basis observables in Fig. \ref{fig:wp_l_scan} is a bit worse than for the mass basis observables, but is still quite good for larger S/B.
It is notable that the results from the Lund basis observables here are much more uniform across the $(\rho,\Sigma)$ landscape than they are for the mass basis observables.
This could be due to the fact that the relevant features in the mass basis are much finer than they are in the Lund basis, as can be seen by comparing Fig. \ref{fig:wp_m_topics} and Fig. \ref{fig:wp_l_topics}.
This is partially due to the logarithmic binning used in the Lund basis observables, but is mostly due to the actual choice of observables.
In Fig. \ref{fig:wp_l_theme_extracted} we show an example of the themes extracted using the VI algorithm for these scans, with S/B$=5\%$, using Lund basis observables, and with $(\rho,\Sigma)=(0.1,1.0)$ yielding the largest inverse perplexity in the scan.

\clearpage
\clearpage

We see that the typical signal features are well distinguishable in one of the themes (theme 2). 
The two clusters in the leading jet distribution of this theme correspond to the  decays of the $\phi$ boson and the $W$ boson, with the single cluster in the subleading jet corresponding to the  decay of the $W$ boson.
As in the mass basis example for boosted $t\bar t$, we observe some notable differences when comparing the two themes to pure signal and background distributions, especially in the soft ($k_T\to 0$) and collinear ($\Delta \to 0$) regions of the Lund plane. Theme 2 predominantly captures that hard splittings associated with the massive resonance decays, while the softer splittings are predominantly captured by theme 1. 
Interestingly, it seems that the algorithm in this case picked up some distinguishable features of the signal (a deficit below the $W$ peak) even in the non-perturbative (low $k_T$) regime. We warn however that these effects are very subtle and the least robust, since they vary considerably dependent on the model priors.

%%%%%%%%%%%%%%%%%%%%%%%%%%%%%%%%%%%%%%%%%%%%%%%%%%
\subsection{Systematics}\label{sec:appsys}
%%%%%%%%%%%%%%%%%%%%%%%%%%%%%%%%%%%%%%%%%%%%%%%%%%

\noindent In order to use the techniques presented above in practice it is important that the VI algorithm produces results which are stable under changes in the random initialisations of the model variables, i.e. the random seed.
Also important is to verify that the algorithm parameters chosen for the inference procedure (see Sec.~\ref{sec:approximateInference}) are sensible given the datasets being used.
The most important algorithm parameters here are the offset $(\tau_0)$, the chunk size ($n_c$), and the number of passes ($n_p$).
The offset affects the learning rate, both the overall magnitude and as a function of the global updates, see Eq.~\eqref{eq:learn}.
The chunk size changes how many events are used to optimise the local parameters before an update on the global parameters is performed.
Finally, the number of passes must simply be large enough such that the algorithm converges.

\subsubsection{Offset}
%%%%%%%%%%%%%%%%%%%%%%%%%%%%%%%%%%%%%%%%%%%%%%%%%%

\begin{figure}[t]
  \centerline{\includegraphics[scale=0.5]{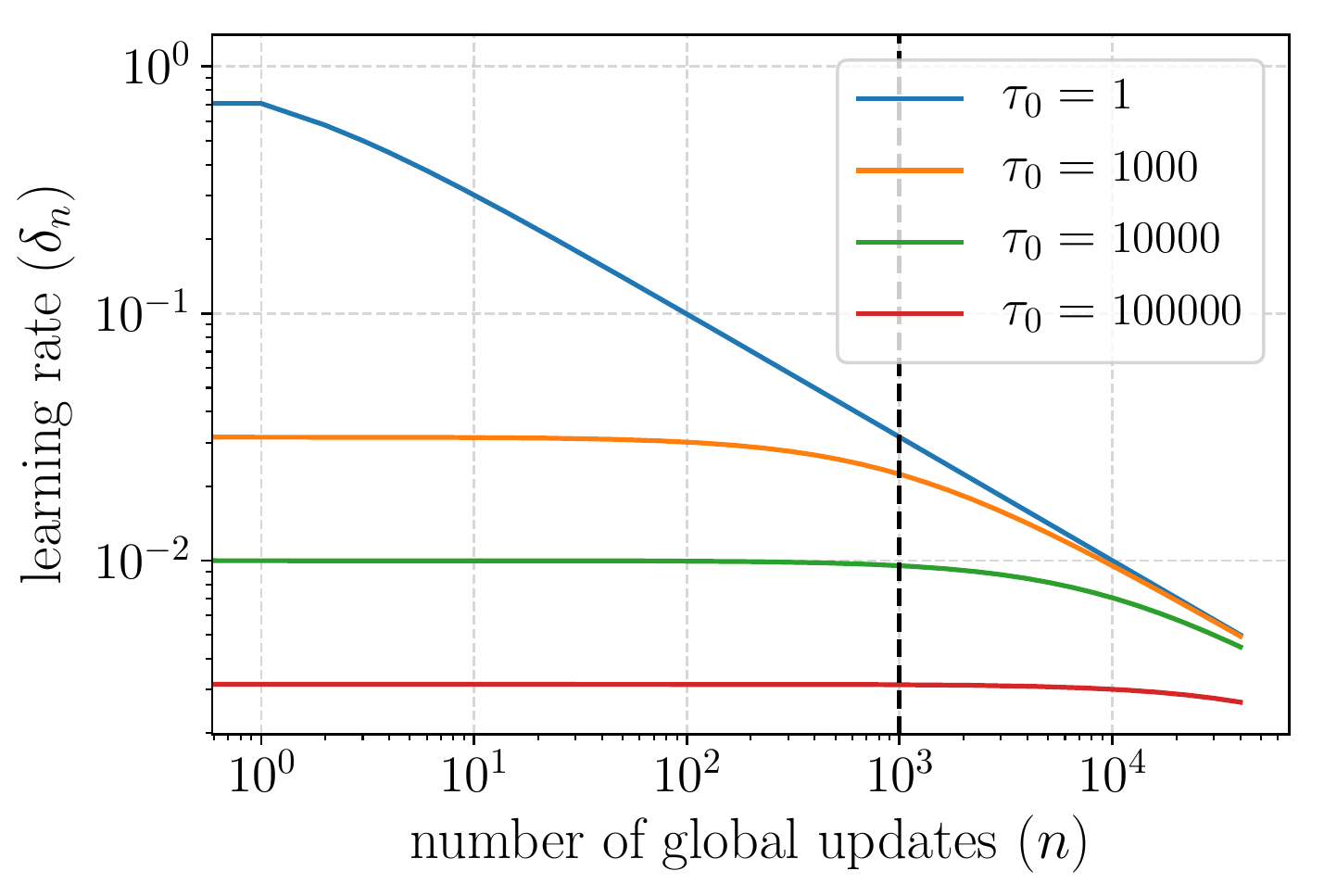}
  }
  \caption{The learning rate as a function of the number of global updates for different offsets, $\tau_0 = 1$-$10^5$.
  The black dashed line indicates the learning rate after $100$ passes when we have $10^5$ events in the sample and a chunk size of $10^4$, as we had in the prior scans in Secs~\ref{sec:ttscans} and~\ref{sec:wpscans}.
   \label{fig:offsetlr}}
\end{figure}

\noindent
We start with the offset, and to be clear what the actual consequences of particular offset values are in the inference algorithm, we show in Fig. \ref{fig:offsetlr} how the offset affects the learning rate ($\delta_n$) as a function of the number of global updates.
We see that the larger offsets inevitably mean a smaller learning rate, but also a learning rate which is more constant across the global updates.
Smaller offsets lead to very large learning rates at earlier global updates, and larger learning rates overall.

\begin{figure}[t]
  \centerline{\includegraphics[scale=0.4]{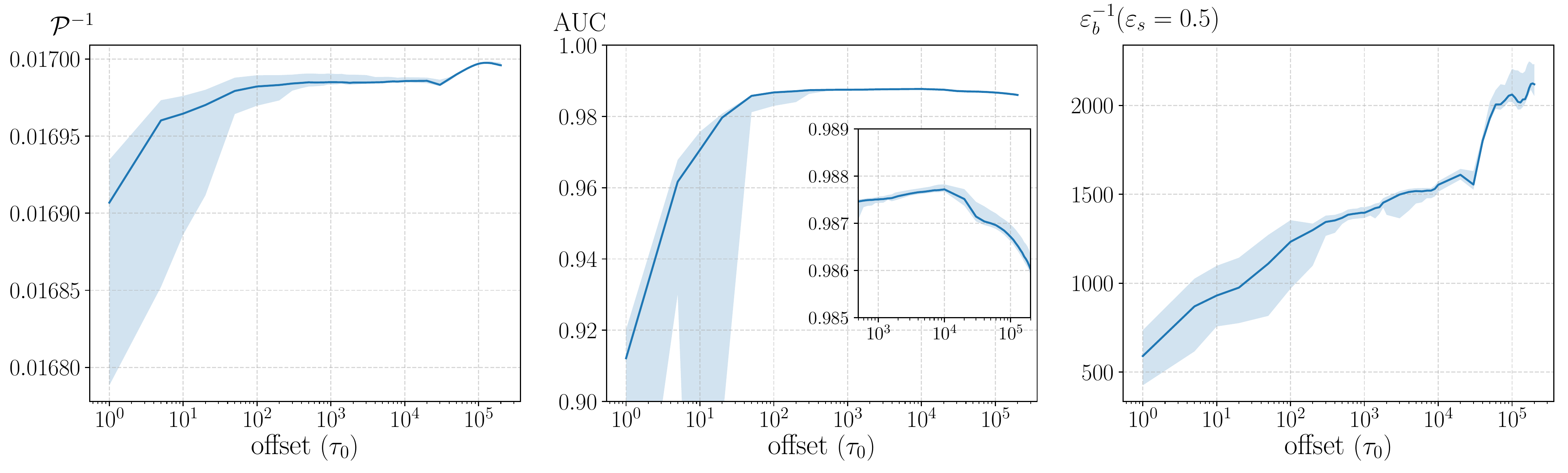}
  }
  \caption{The inverse perplexity (left), AUC (center), and inverse mistag rate (right) as functions of the offset for an event sample with $10^5$ events and a chunk size of $10^4$.
  The calculations were done for $100$ different random seeds, the blue lines show the mean of these and the shaded regions cover the upper and lower standard deviations.
  Separate upper and lower standard deviations are used to show how the actual variances in the performance statistics are typically skewed heavily towards the negative side.
    \label{fig:offsetbig}}
\end{figure}

To demonstrate the effect that the learning rate and offset have on the results, we have chosen a single parameter point from the scans performed on the $W'$/QCD mixed event sample, with $S/B=2.5\%$ and $[\rho,\Sigma]=[0.05,1.3]$ in the mass basis.
We keep all of the parameters as they were in the scan, except now the offset is varied from $1$ to $2\!\times\!10^5$.
For each offset we calculate the perplexity, the AUC, and the inverse mistag rate so that we can analyse changes in performance.
To assess the stability of the algorithm as a function of offset we repeat this for $100$ different random seeds, calculating the mean and the (upper and lower) standard deviation of the resulting distribution.
These results are shown in Fig.~\ref{fig:offsetbig}.

The first clear effect we see is that both the perplexity and the performance of these models increase with the size of the offset, degrading heavily at low offsets.
The reason for this is simple, the learning rate is too large to sufficiently resolve the maxima in the ELBO.
We also see that the random seed induced variance of the results increases considerably at low offsets.
This is partially due to the overall size of the learning rate, but is also affected by the significantly increased learning rate in the initial global updates, as can be seen in Fig.~\ref{fig:offsetlr}.
Because the chunks of data are sampled randomly at the beginning of the analysis, a different random seed means that a different subset of the data will have more influence on the inference, hence the larger variance.
The second effect we see is the change in behaviour at very large offset.
The AUC and inverse mistag are both good measures of performance for the model so we might expect that an increase in one leads to an increase in the other, however we see here that this is not the case.
At offset $\sim 10^4$ the AUC begins to degrade while the inverse mistag at fixed efficiency of $\varepsilon_s = 0.5$ continues to improve somewhat.

\begin{figure}[t]
  \centerline{\includegraphics[scale=0.5]{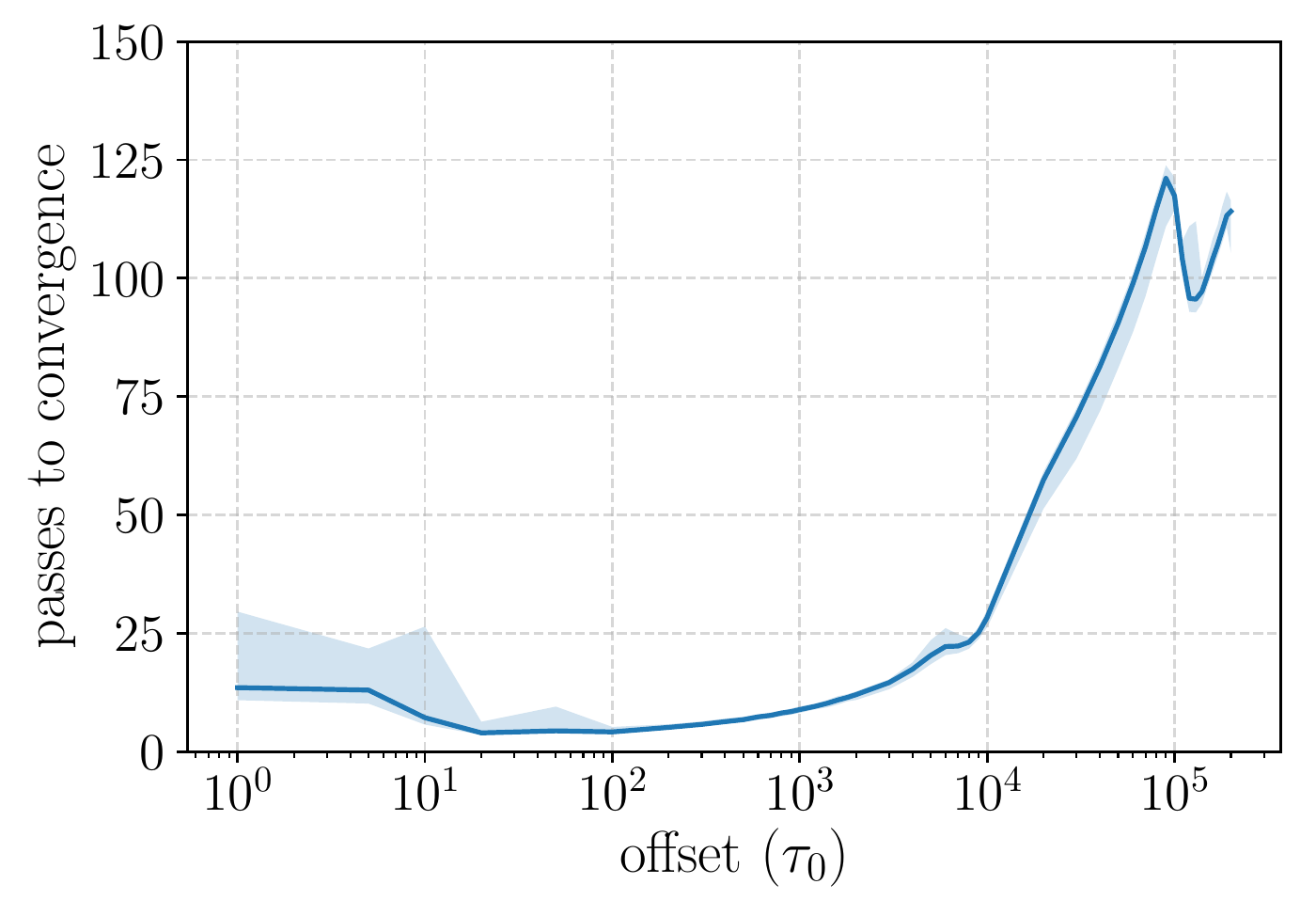}
  }
  \caption{Number of passes needed for convergence as a function of the offset, for an event sample with $10^5$ events and a chunk size of $10^4$.
  The calculations were done for $100$ different random seeds, the blue lines show the mean of these and the shaded regions cover the upper and lower standard deviations.
  Separate upper and lower standard deviations are used to show how the actual variances in the number of required passes are skewed with respect to the mean.   \label{fig:offsetbigconv}}
\end{figure}

The learning rate also affects the speed of convergence of VI. In the algorithm described in Sec.~\ref{sec:approximateInference} we allow the algorithm to run for a fixed number of passes over the data without checking for convergence. However one could easily change this to check explicitly for convergence and end the algorithm early. In  Fig.~\ref{fig:offsetbigconv} we look at how many passes over the data the algorithm takes to converge, seeing that runs with larger offsets take much longer to converge.
This is easily understood due to the smaller learning rate implied by larger offsets.

From these observations we deduce that an offset in the range $10^3$-$10^4$ is the best choice for both the performance and stability of the inference algorithm as applied to our example datasets. Correspondingly, the suitability of a paticular offset choice on other datasets can be readily verified by checking for convergence as well as model perplexity dependence on this algorithm parameter.

\subsubsection{Chunk size}
%%%%%%%%%%%%%%%%%%%%%%%%%%%%%%%%%%%%%%%%%%%%%%%%%%

\begin{figure}[t]
  \centerline{\includegraphics[scale=0.4]{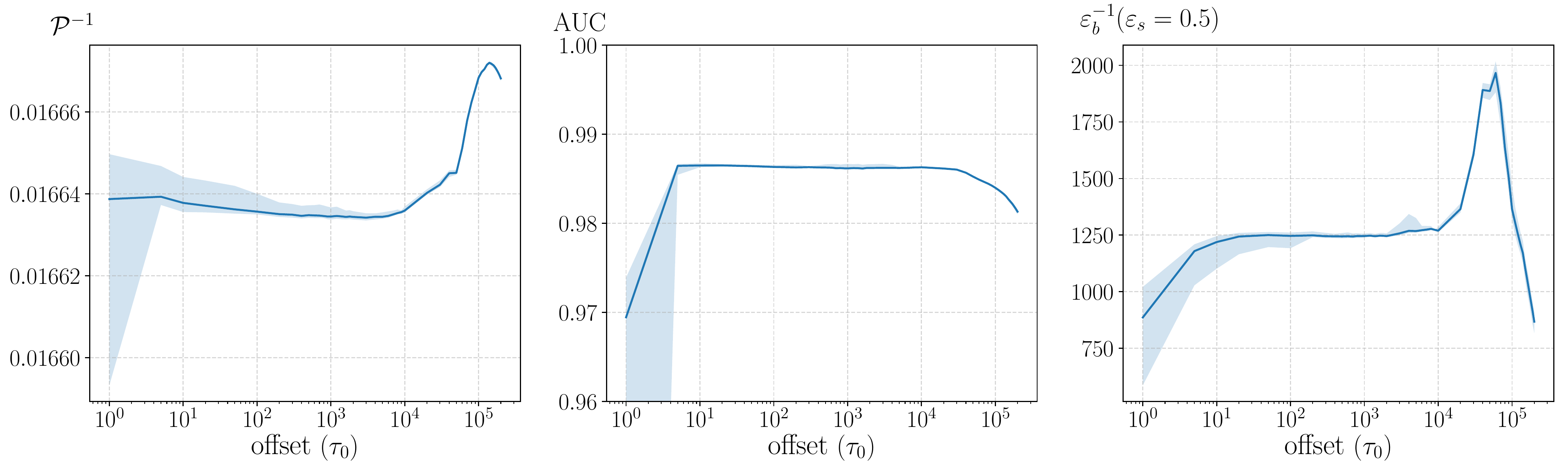}
  }
  \caption{The inverse perplexity (left), AUC (center), and inverse mistag rate (right) as functions of the offset for an event sample with $10^4$ events and a chunk size equal to the size of the event sample.
  The calculations were done for $100$ different random seeds, the blue lines show the mean of these and the shaded regions cover the upper and lower standard deviations.
  Separate upper and lower standard deviations are used to show how the actual variances in the performance statistics are typically skewed heavily towards the negative side.   \label{fig:offsetsmall}}
\end{figure}

\noindent
In the prior scans in Secs.~\ref{sec:ttscans} and~\ref{sec:wpscans}  the chunk size was $10^4$ while the samples contained almost $10^5$ events, i.e. $10$ chunks per pass over the sample.
Since we are looking for rare signals it is possible that the signal events could be very unevenly distributed throughout these different chunks, resulting in each chunk having significantly different perplexity, i.e. ELBO.
Therefore the algorithm would essentially be attempting to optimise one model for these $10$ different chunks, and the resulting posterior approximation would fail to accurately describe the true posterior.
To test that this is not an issue in the scans, we have performed the same offset scan as in Fig.~\ref{fig:offsetbig} but now for a smaller event sample ($10^4$ events), where the global updates are performed only after seeing the whole dataset, i.e. the chunk size is equal to the size of the event sample.
We see these results in Fig.~\ref{fig:offsetsmall} and it is clear that while they differ slightly, qualitatively the same behavior is observed in the perplexity and performance at different offsets.

\begin{figure}[t]
  \centerline{\includegraphics[scale=0.4]{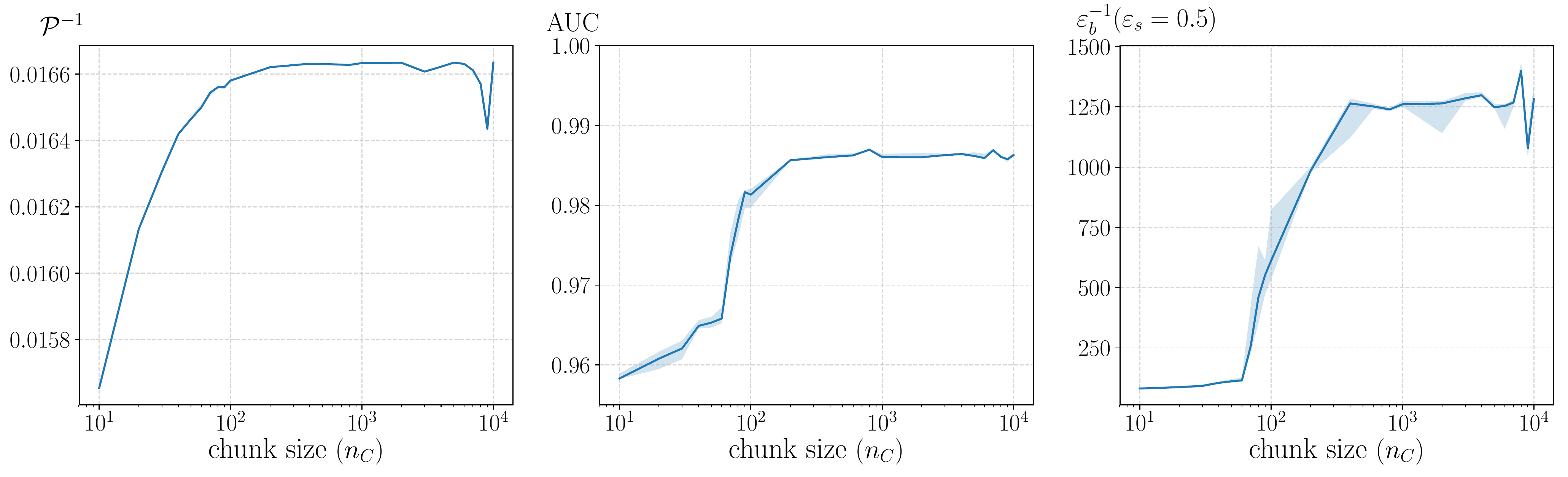}
  }
  \caption{The inverse perplexity (left), AUC (center), and inverse mistag rate (right) as functions of the chunk size for a mixed $W'$/QCD event sample with $10^4$ events and a chunk size equal to the size of the event sample.
  The offset is also varied such that the learning rate is held constant.
  The calculations were done for $100$ different random seeds, the blue lines show the mean of these and the shaded regions cover the upper and lower standard deviations.
  Separate upper and lower standard deviations are used to show how the actual variances in the performance statistics are typically skewed heavily towards the negative side.   \label{fig:chunksizesmall}}
\end{figure}

To properly study the effect of changing the chunk size we need to find a better way to compare models trained with different chunk sizes.
Changing the chunk size significantly affects how much of the data the algorithm analyses before it converges, and we would like to disentangle this effect from the effect due to less data being analysed per global update in the algorithm.
To do this we vary the offset simultaneously with the chunk size such that the learning rate at one pass over the data is held constant.
The example we use is again the $W'$/QCD mixed sample of a total $10^4$ events, with $S/B=2.5\%$ and $[\rho,\Sigma]=[0.05,1.3]$ in the mass basis.
The learning rate is held constant to what it would be if we had an offset of $10^3$ and a chunk size also equal to $10^4$ events.
The chunk size is varied from $10$ up to $10^4$, meaning that the offset varies from $1$ to $1000$.
The results are shown in Fig. \ref{fig:chunksizesmall}, where we clearly see the disadvantages in using very small chunk sizes.
When the chunk size reaches $\mathcal{O}(5\%)$ of the size of the event sample the perplexity and performance statistics reach a plateau.
As we vary the chunk size we see that the results are not very sensitive to the random seed. This is because the learning rate is kept at a constant (small enough) value by also varying the offset accordingly. 

So while choosing the chunk size to be equal to the size of the event sample is certainly a good idea, especially with smaller datasets and rare signals, we conclude that for the event samples that we have analysed in this paper setting the chunk size to only a fraction of total dataset does not significantly impede the quality or robustness of VI while significantly improving its convergence.
In particular, our reasoning in choosing the chunk size to be $10^4$ rather than $10^5$ for our prior scans in Secs.~\ref{sec:ttscans} and~\ref{sec:wpscans} is thus simply that the algorithm converges $10$ times faster.

%%%%%%%%%%%%%%%%%%%%%%%%%%%%%%%%%%%%%%%%%%%%%%%%%%
%
\section{Conclusions}\label{sec:conclusions}
%
%%%%%%%%%%%%%%%%%%%%%%%%%%%%%%%%%%%%%%%%%%%%%%%%%%

\noindent
In this work we have described a general unsupervised framework capable of learning rare patterns in event data collected at high-energy colliders.
We use a Bayesian probabilistic modelling technique called Latent Dirichlet Allocation (LDA), an unsupervised ML approach that was first introduced in the context of BSM collider physics in a previous paper \cite{Dillon:2019cqt}. 
We started by representing individual collider events as sequences of binned exchangeable measurements, and assumed a simplified picture in which the events are generated by sampling these measurements from some underlying joint probability distribution.
The assumption of exchangeability of measurements in an event guarantees, through de Finnetti's theorem, that the sequence of measurements in an event are conditionally dependent on a latent variable sampled from (marginalised over) a prior distribution in latent space. 
Through some basic assumptions on this latent space we arrived at the LDA model, which we focus on throughout the paper.
LDA is a mixed-membership model, meaning that under this model it is assumed that the measurements in each event are assumed to have been sampled from multiple (two, in our case) different multinomial distributions --  {\it themes}. These themes encode information on the underlying structure, i.e. hidden patterns, in the event data represented in terms of binned measurements.
The mixing proportions of themes are sampled from a prior taking the form of a Dirichlet distribution, a parametric family of distributions over the simplex. 
Mixed membership models have the advantage of describing different events which share features arising from the same underlying physical source.
Depending on the Dirichlet prior, the generative model can naturally describe event samples where certain combinations of measurements appear rarely, which is crucial for uncovering rare signals.
Given the LDA model and the event data, we described in detail a stochastic variational inference technique for approximating or learning the underlying themes from which the data is assumed to have been generated.
We then described how the extracted themes can be used to construct a classifier to cluster events into two categories, potentially aligned with the background and signal classes.
We finally identified a measure of classification performance based solely the learned themes, {\it the perplexity}, which does not require truth labels to compute and can thus be extracted directly from mixed data. In particular, we found that perplexity correlates strongly with the widely used traditional measures of classification performance based on the ROC curves -- the AUC and inverse-mistag rate at fixed efficiency, which do require truth labels.
%We also described in detail the training procedure for LDA models on data using an approximate technique known as (stochastic) variational inference (implemented in the {\tt Gensim} package used throughout this work). \Dario{[Maybe Barry can fill in here :)]}.

To demonstrate the power of this technique we considered the analysis of di-jet events at the LHC focusing on two benchmark examples; boosted SM $t\bar t$ production and a hypothetical BSM production of $W^\prime \to (\phi \to WW) W$. We described in detail how to pre-process the event data to express each event as a sequence of exchangeable measurements, and how the generative model for di-jet events is to be interpreted using LDA.
Our choice of jet substructure observables that we used in the analysis is based upon high level observable combinations that have previously been shown to be good for identifying massive resonance decay chains within large radius jets with supervised methods: the traditional mass drop basis (see e.g. Ref.~\cite{Kaplan:2008ie}) and the primary Lund plane basis \cite{Dreyer:2018nbf}.\footnote{We note in passing the in principle LDA can  be trained on any general combination of high-level observables used in supervised classification that has significant discriminating power, thus in principle allowing to promote supervised classifiers to unsupervised ones, given enough measurement co-occurances in the data.}  Through a study of the classification power of these different observables, and of how strong their co-occurrences are in the data, we have identified most promising pairs of observables in each basis for our unsupervised classification approach.

The results for each of the benchmark di-jet examples from this study are presented in Sec.~\ref{sec:ldaresults}.
Using the perplexity, AUC, and inverse-mistag rate at a fixed signal efficiency as performance indicators, we analysed how well the two-theme LDA models classified events over a large range of values of Dirichlet prior parameters ($(\rho, \Sigma)$).
For each benchmark we considered six different samples with varying S/B, ranging from $0.01$ to $1.0$ for the boosted top-quark example, and from $0.005$ to $0.1$ for the $W'$ example, including background only samples for reference.
For both benchmarks the mass drop observables generally outperform the Lund observables in classification, however both choices lead to complementary results with the extracted themes in each case holding valuable information about the signal and background processes.
From the results it is clear that the inference algorithm was able to separate measurement patterns corresponding to the massive resonance decays within the signal jets from patterns corresponding to light QCD emissions present within all jets.
This is achieved due to the mixed-membership nature of the generative model, where QCD-like patterns found both in the signal and background jets were identified as having been sampled from the same theme describing QCD-like splittings in the jet substructure.
%Using the Lund basis we have seen that for mixed samples containing the $W'$ signal, the inference algorithm was able to separate the hard splittings corresponding to massive resonance decays within the signal jets from the non-perturbative splittings at low $k_T$.

Finally, in Sec.~\ref{sec:appsys} we studied how the results and performance of the chosen inference technique depend on the tunable parameters of the algorithm, in particular the chunk size and the offset. 
We demonstrated that the results of the algorithm are in fact stable over a large range of these parameters, and that the algorithm tends to converge within $\lesssim 100$ passes for the example datasets.

Perhaps the most important result of this work is that over the $(\rho,\Sigma)$ Dirichlet parameter plane the AUC and inverse-mistag rate, calculated using truth label information, are strongly correlated with the perplexity, which is calculated without truth label information. This implies that, not only can perplexity be used as a practical measure to asses LDA model convergence, but it can also provide guidance when selecting the most viable and robust Dirichlet priors for unsupervised collider analyses and searches.
By allowing the algorithm to select optimal $\rho$ and $\Sigma$ parameters we would not need to perform a search for each choice of parameters considered, meaning that there would be no contribution to the trials factor due to these parameters.
This result is a crucial step towards the next part of this work programme, constructing a full unsupervised di-jet search strategy for new physics at the LHC using LDA.
In a recent letter \cite{Aad:2020cws} the ATLAS collaboration published an analysis of a weakly supervised di-jet resonance search in which contributions to the trials factor associated to the masses of the final state jets are eliminated by allowing the ML algorithm to define the classifier using the event data alone.
We would like to stress that the method presented in this paper also benefits from such a reduction in the trials factor.
In fact, because we represent each jet as a sequence of splittings corresponding to possibly many massive resonance intermediate decays, this method has the potential to describe arbitrarily complicated jet substructure signatures without paying any penalty in the trials factor.
We reserve a full discussion of the search strategy to an upcoming publication. 

\acknowledgments
\noindent
We would like to thank Bryan Zaldivar, Ezequiel Alvarez, and especially Jesse Thaler for several enlightening discussions. We also thank Jack Collins for generously providing the $W'-\phi$ NP model implementation for use in aMC@NLO. JFK deeply appreciates the continuing hospitality and computing support of CERN without which this project would not have been possible. MS would like to thank the Jozef Stefan Institute for its enormous hospitality. BD and JFK acknowledge the financial support from the Slovenian Research Agency (research core funding No. P1-0035 and J1-8137). DAF is supported by the Swiss National Science Foundation (SNF) under contract 200021-159720. This article is based upon work from COST Action CA16201 PARTICLEFACE supported by COST (European Cooperation in Science and Technology).

\bibliographystyle{JHEP}
\bibliography{current}

\end{document}